\documentclass[12pt]{iopart}

\usepackage{iopams}  
\usepackage{graphicx}
\begin{document}

\title[Fermi surface pockets in electron-doped iron superconductor]
{Fermi surface pockets in electron-doped iron superconductor by Lifshitz transition}

\author{Jose P. Rodriguez \& Ronald Melendrez}

\address{Department of Physics and Astronomy, California State University at Los Angeles, Los Angeles, USA}
\ead{jrodrig@calstatela.edu}
\vspace{10pt}
\begin{indented}
\item[]June 2018
\end{indented}

\begin{abstract}
The Fermi surface pockets that lie at the corner of the two-iron Brillouin zone
in heavily electron-doped iron selenide superconductors are accounted for by an
extended Hubbard model over the square lattice of iron atoms that 
includes the principal $3d_{xz}$ and $3d_{yz}$ orbitals.
At half filling, and in the absence of intra-orbital next-nearest neighbor hopping,
perfect nesting between electron-type and hole-type Fermi surfaces
at the the center and at the corner of the one-iron Brillouin zone is revealed.
It results in hidden magnetic order in the presence of magnetic frustration
within mean field theory.  An Eliashberg-type calculation that includes spin-fluctuation exchange
finds that the Fermi surfaces undergo a Lifshitz transition to electron/hole
Fermi surface pockets centered at the corner of the two-iron Brillouin zone
as on-site repulsion grows strong.  In agreement with 
angle-resolved photoemission spectroscopy on iron selenide high-temperature superconductors,
only the two electron-type Fermi surface pockets
remain after a rigid shift in energy of the renormalized band structure
by strong enough electron doping.  At the limit of strong on-site repulsion,
a spin-wave analysis of the hidden-magnetic-order state finds a ``floating ring''
of low-energy spin excitations centered at the checkerboard wavenumber $(\pi,\pi)$.
This prediction compares favorably with recent observations of low-energy
spin resonances around $(\pi,\pi)$ in intercalated iron selenide by inelastic neutron scattering.
\end{abstract}

%
%
%
\maketitle
%
%

\section{Introduction}
The family of iron-based superconductors first discovered ten years ago has
established a new route to high critical temperatures\cite{hosono_08}.  
In particular,
stoichiometric FeSe becomes superconducting below 
a modest critical temperature of $8$ K.
Electron doping of FeSe  raises the critical temperature
dramatically into the range $40$-$110$ K, however\cite{zhang_14,ge_15}.
The latter has been achieved by laying
a monolayer of FeSe over a substrate\cite{xue_12,liu_12,zhou_13,fan_15},
by intercalating layers of FeSe with organic compounds\cite{zhao_16,niu_15,yan_16},
by dosing thin films of FeSe with alkali metals\cite{miyata_15,wen_16,song_16},
and by applying a gate voltage to thin films of FeSe\cite{lei_16,hosono_16}.
Angle-resolved photoemission spectroscopy (ARPES)
on such heavily electron-doped FeSe
reveals two electron-type Fermi surface
pockets at the corner of the two-iron Brillouin zone\cite{liu_12,zhou_13}.
It also reveals
buried hole bands at the center of the two-iron Brillouin zone that
lie $60$ meV below the bottom of the former electron bands\cite{zhao_16,lee_14}.
At low temperature, ARPES also finds an isotropic gap 
at the electron Fermi surface pockets
consistent with a superconducting state\cite{zhao_16,peng_14}.
The energy gap at the Fermi surface is confirmed by scanning tunneling microscopy
on heavily electron-doped FeSe\cite{fan_15,yan_16,song_16}.
By comparison with stoichiometric FeSe,
which has a relatively low critical temperature,
it has been argued that the phenomenon of  high-temperature superconductivity
in heavily electron-doped FeSe is due to 
the appearance of a new electronic groundstate\cite{jpr_17,huang_hoffman_16}.

In addition, recent inelastic neutron scattering experiments on intercalated FeSe
find a ring of low-energy magnetic excitations centered at the 
wave number $(\pi/a,\pi/a)$ associated with N\'eel order over the square
lattice of iron atoms in a single layer\cite{davies_16,pan_17,ma_17}. 
Here, $a$ denotes the lattice constant.
It has been suggested recently by one of the authors that this ring of low-energy
magnetic excitations is a result of hidden N\'eel order among the
iron $3d$ orbitals\cite{jpr_17,jpr_16}.  (See Fig. \ref{fltng_rng}.)
Such hidden magnetic order can emerge because of frustration among local
magnetic moments over
the square lattice of iron atoms in FeSe\cite{jpr_ehr_09,jpr_10}. 
One of the authors has shown that adding
electrons to the local magnetic moments at half filling
results in electron-type Fermi surface pockets at 
the corner of the two-iron Brillouin zone,
but with no Fermi surface at the center\cite{jpr_17}.
It is important to mention, at this stage, that conventional band structure 
calculations for electron-doped FeSe typically predict
additional hole-type Fermi surface pockets
at the center of the two-iron Brillouin zone,
in marked contrast to ARPES measurements\cite{zhao_16,peng_14,B&C}

In this paper, we shall introduce an extended Hubbard model over the square lattice
of iron atoms in heavily electron-doped FeSe that harbors hidden N\'eel order
among the $3d_{xz}/3d_{yz}$ orbitals because of perfect nesting of
electron-type and of hole-type Fermi surfaces 
at the  center and at the corner of the one-iron Brillouin zone,
with nesting wavenumber $(\pi/a,\pi/a)$.  
True N\'eel order is suppressed because of
magnetic frustration from super-exchange interactions across the Se atoms\cite{jpr_ehr_09,jpr_10}.
At half filling,
mean field theory similar to that for the one-orbital 
Hubbard model over the square lattice\cite{hirsch_85}
finds a stable hidden spin-density wave (hSDW) at the same wavenumber,
but with a nodal $D_{xy}$ gap in the quasi-particle spectrum at the Fermi surface.
Also in analogy with the one-orbital Hubbard model over the square
lattice\cite{schrieffer_wen_zhang_89,singh_tesanovic_90,chubokov_frenkel_92},
we identify two collective modes of the mean field theory
that represent spin-wave excitations of the hSDW.
They vanish in energy at the N\'eel wave number $(\pi/a,\pi/a)$ 
with an acoustic dispersion.
We argue in the Discussion section that degeneracy of these hidden Goldstone modes
with true spin excitations results in a ring of low-energy magnetic excitations
similar to what is observed 
by inelastic neutron scattering in intercalated FeSe\cite{davies_16,pan_17,ma_17}.

Next, we shall formulate an Eliashberg theory\cite{eliashberg_60,eliashberg_61,schrieffer_64} 
for the extended Hubbard model of 
a single layer of heavily electron-doped FeSe
that is  based on exchange of the above hidden spin-wave excitations
by electrons and by holes (particle-hole channel).
A solution of the associated Eliashberg equations\cite{scalapino_69}
finds a Lifshitz transition of the electron/hole Fermi surfaces
to pockets centered at the corner of the two-iron Brillouin zone
at moderate to strong on-site Coulomb repulsion.
This result is consistent with similar results obtained by one of the authors
at the limit of strong on-site Coulomb repulsion\cite{jpr_17},
which predict electron-type Fermi surface pockets alone
at the corner of the two-iron Brillouin zone at {\it any} level of electron doping.
At strong but finite on-site Coulomb repulsion,
the present Eliashberg-type calculation
finds a threshold electron doping
beyond which electron-type Fermi surface pockets appear alone.
Below, we introduce the two-orbital Hubbard model
for heavily electron-doped FeSe.

\section{Perfect nesting of Fermi surfaces}
We retain only the $3d_{xz}/3d_{yz}$ orbitals of the iron atoms 
in the following description
for a single layer of heavily electron-doped FeSe. In particular,
let us work in the isotropic basis of orbitals
$d-=(d_{xz}-id_{yz})/{\sqrt 2}$ and $d+=(d_{xz}+id_{yz})/{\sqrt 2}$.
The kinetic energy is governed by the hopping Hamiltonian
\begin{equation}
H_{\rm hop} =  
-\sum_{\langle i,j \rangle} (t_1^{\alpha,\beta} c_{i, \alpha,s}^{\dagger} c_{j,\beta,s} + {\rm h.c.}) 
-\sum_{\langle\langle  i,j \rangle\rangle} (t_2^{\alpha,\beta} c_{i, \alpha,s}^{\dagger} c_{j,\beta,s} + {\rm h.c.}),
\label{hop}
\end{equation}
where  repeated indices $\alpha$ and $\beta$ are summed over the $d-$ and $d+$ orbitals,
where repeated index $s$ sums over electron spin,
and where $\langle i,j\rangle$ and $\langle\langle i,j\rangle\rangle$
represent nearest neighbor (1) and next-nearest neighbor (2) links on the
square lattice of iron atoms.
Above, $c_{i, \alpha,s}$ and $c_{i, \alpha,s}^{\dagger}$
denote annihilation and creation operators for an
electron of spin $s$ in orbital $\alpha$ at site $i$.
The reflection symmetries shown by a single layer of FeSe imply
that the above intra-orbital and inter-orbital hopping matrix elements
show $s$-wave and $d$-wave symmetry, respectively\cite{raghu_08,Lee_Wen_08,jpr_mana_pds_14}.  
In particular,
nearest neighbor hopping matrix elements satisfy
\begin{eqnarray}
t_1^{\pm \pm} ({\bf {\hat x}}) &=& t_1^{\parallel} = t_1^{\pm \pm} ({\bf {\hat y}}) \nonumber\\
t_1^{\pm\mp} ({\bf {\hat x}}) &=& t_1^{\perp} = -t_1^{\pm \mp} ({\bf {\hat y}}),
\label{t1}
\end{eqnarray}
with real $t_1^{\parallel}$ and $t_1^{\perp}$,
while next-nearest neighbor hopping matrix elements satisfy
\begin{eqnarray}
t_2^{\pm \pm} ({\bf {\hat x}}+{\bf {\hat y}}) = \; t_2^{\parallel} &=& t_2^{\pm \pm} ({\bf {\hat y}}-{\bf {\hat x}}) \nonumber\\
t_2^{\pm \mp} ({\bf {\hat x}}+{\bf {\hat y}}) = \pm t_2^{\perp} &=& -t_2^{\pm \mp} ({\bf {\hat y}}-{\bf {\hat x}}),
\label{t2}
\end{eqnarray}
with real $t_2^{\parallel}$ and pure-imaginary $t_2^{\perp}$.

The above hopping Hamiltonian is easily diagonalized by 
plane waves of $d_{x(\delta)z}$ and $i d_{y(\delta)z}$ orbitals that are rotated
with respect to the principal axis by a phase shift $\delta({\bi k})$:
\begin{eqnarray}
|{\bi k}, d_{x(\delta)z}\rangle\rangle &=&
{\cal N}^{-1/2} \sum_i e^{i{\bi k}\cdot{\bi r}_i} 
[e^{i\delta({\bi k})} |i, d+\rangle + e^{-i\delta({\bi k})} |i, d-\rangle], \nonumber\\
i|{\bi k}, d_{y(\delta)z}\rangle\rangle &=&
{\cal N}^{-1/2} \sum_i e^{i{\bi k}\cdot{\bi r}_i} 
[e^{i\delta({\bi k})} |i, d+\rangle - e^{-i\delta({\bi k})} |i, d-\rangle],
\label{plane_waves}
\end{eqnarray}
where ${\cal N} = 2 N_{\rm Fe}$ is the number of iron site-orbitals.
Their energy eigenvalues are respectively given by
$\varepsilon_+({\bi k}) = \varepsilon_{\parallel}({\bi k}) + |\varepsilon_{\perp}({\bi k})|$ and
$\varepsilon_-({\bi k}) = \varepsilon_{\parallel}({\bi k}) - |\varepsilon_{\perp}({\bi k})|$,
where
\numparts
\begin{eqnarray}
\label{mtrx_lmnt_a}
\varepsilon_{\parallel}({\bi k}) &=& -2 t_1^{\parallel} (\cos k_x a + \cos k_y a)
-2 t_2^{\parallel} (\cos k_+ a + \cos k_- a) \\
\label{mtrx_lmnt_b}
\varepsilon_{\perp}({\bi k}) &=& -2 t_1^{\perp} (\cos k_x a - \cos k_y a)
-2 t_2^{\perp} (\cos k_+ a - \cos k_- a)
\end{eqnarray}
\endnumparts
are diagonal and off-diagonal matrix elements,
with  $k_{\pm} = k_x \pm k_y$.
The phase shift $\delta({\bi k})$ is set by
$\varepsilon_{\perp}({\bi k}) = |\varepsilon_{\perp}({\bi k})| e^{i 2 \delta({\bi k})}$.
Specifically,
\numparts
\begin{eqnarray}
\label{c_2dlt}
\cos\,2\delta({\bi k}) &=& {-t_1^{\perp}(\cos\, k_x a - \cos\, k_y a)\over
{\sqrt{t_1^{\perp 2}(\cos\, k_x a - \cos\, k_y a)^2 +
|2 t_2^{\perp}|^2 (\sin\, k_x a)^2 (\sin\, k_y a)^2}}}, \\
\label{s_2dlt}
\sin\,2\delta({\bi k}) &=& {2 (t_2^{\perp} / i)(\sin\, k_x a) (\sin\, k_y a)\over
{\sqrt{t_1^{\perp 2}(\cos\, k_x a - \cos\, k_y a)^2 +
|2 t_2^{\perp}|^2 (\sin\, k_x a)^2 (\sin\, k_y a)^2}}}. 
\end{eqnarray}
\endnumparts
It is notably singular at ${\bi k} = 0$ and ${\bi Q}_{\rm AF}$,
where the matrix element $\varepsilon_{\perp}({\bi k})$ vanishes.

\begin{figure}
\hspace*{2cm}\includegraphics[scale=1.00]{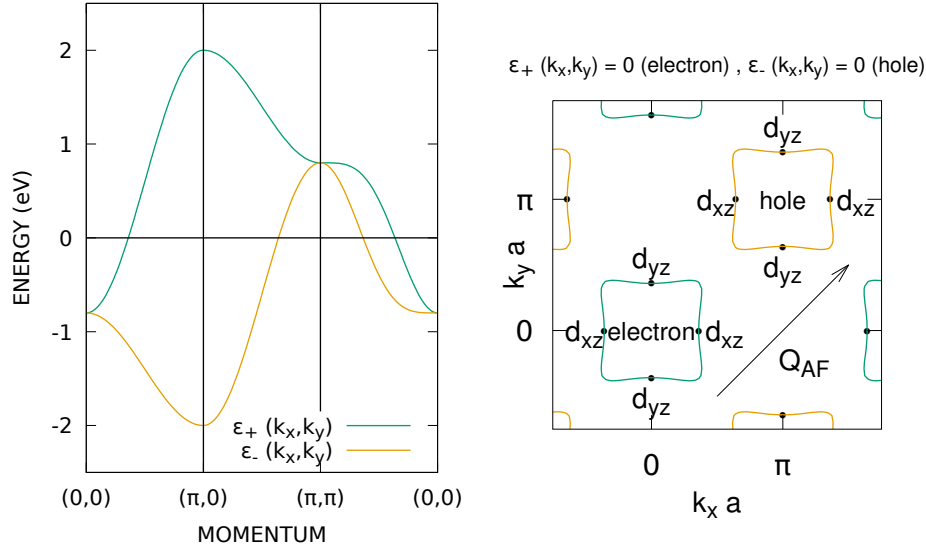}
\caption{Band structure with perfectly nested Fermi surfaces at half filling,
$\varepsilon_+({\bi k}) = 0$ and  $\varepsilon_-({\bi k}) = 0$,
with hopping matrix elements
$t_1^{\parallel} = 200$ meV, $t_1^{\perp} = 500$ meV, $t_2^{\parallel} = 0$,
and $t_2^{\perp} = 100\, i$ meV. Point nodes of quasi-particle gap are marked on Fermi surfaces.}
\label{FS0}
\end{figure}

Let us now turn off next-nearest neighbor intra-orbital hopping: $t_2^{\parallel} = 0$.
Notice, then, that the above energy bands satisfy the perfect nesting condition
\begin{equation}
\varepsilon_{\pm}({\bi k}+{\bi Q}_{\rm AF}) = - \varepsilon_{\mp}({\bi k}),
\label{prfct_nstng}
\end{equation}
where ${\bi Q}_{\rm AF} = (\pi/a,\pi/a)$ is the N\'eel ordering vector on the square
lattice of iron atoms. 
The relationship (\ref{prfct_nstng}) is an expression of a particle-hole symmetry that 
the hopping Hamiltonian (\ref{hop}) exhibits at $t_2^{\parallel} = 0$.
(See \ref{ppndx_p_h}.)
As a result, it can easily be shown that the Fermi level
at half filling of the bands lies at $\epsilon_{\rm F} = 0$.  
Figure \ref{FS0} shows such
perfectly nested electron-type and hole-type Fermi surfaces for hopping 
parameters $t_1^{\parallel} = 200$ meV, $t_1^{\perp} = 500$ meV, $t_2^{\parallel} = 0$
and $t_2^{\perp} =  100\, i$ meV.

We shall now demonstrate how the perfectly nested Fermi surfaces shown
by Fig. \ref{FS0} can result in an instability 
to long-range hidden N\'eel  order.
It is useful to first write the creation operators for 
the eigenstates (\ref{plane_waves}) of the 
electron hopping Hamiltonian, $H_{\rm hop}$:
\begin{equation}
c_s^{\dagger}(n,{\bi k}) = {\cal N}^{-1/2} \sum_i \sum_{\alpha=0,1}
(-1)^{\alpha n} e^{i(2\alpha-1)\delta(\bi k)} e^{i{\bi k}\cdot{\bi r}_i} c_{i,\alpha,s}^{\dagger},
\label{ck}
\end{equation}
where $\alpha = 0$ and $1$ index the $d-$ and $d+$ orbitals, 
and where $n=1$ and $2$ index the anti-bonding and bonding orbitals
$(-i) d_{y(\delta)z}$ and $d_{x(\delta)z}$.
The inverse of the above is then
\begin{equation}
c_{i,\alpha,s}^{\dagger} = {\cal N}^{-1/2} \sum_{\bi k} \sum_{n=1,2}
(-1)^{\alpha n} e^{-i(2\alpha-1)\delta({\bi k})} e^{-i{\bi k}\cdot{\bi r}_i} c_s^{\dagger}(n,{\bi k}).
\label{ci}
\end{equation}
It is then straight-forward to show that the spin magnetization 
for true ($m=0$) or for hidden ($m=1$) magnetic order,
\begin{equation}
S_z (m,{\bi Q}_{\rm AF}) = \sum_i\sum_{\alpha} (-1)^{\alpha m} e^{i{\bi Q}_{\rm AF}\cdot{\bi r}_i} {1\over 2} (n_{i,\alpha,\uparrow} - n_{i,\alpha,\downarrow}),
\label{Sz}
\end{equation}
takes the form
\begin{eqnarray}
S_z (m,{\bi Q}_{\rm AF}) = {1\over 2}\sum_s\sum_{\bi k}\sum_{n,n^{\prime}}
({\rm sgn}\, s) {\cal M}_{n,{\bi k};n^{\prime},{\bar{\bi k}}} c_s^{\dagger}(n^{\prime},{\bar{\bi k}}) c_s(n,{\bi k}),\nonumber\\
\label{sz}
\end{eqnarray}
where ${\bar{\bi k}} = {\bi k} + {\bi Q}_{\rm AF}$.
The above matrix element is computed in \ref{ppndx_a}.
Importantly, it is given by
\begin{equation}
{\cal M}_{n,{\bi k};n^{\prime},{\bar{\bi k}}} =
\cases{
\pm \sin 2\delta({\bi k})  & for $n^{\prime} = n + m\; ({\rm mod}\; 2)$, \\
\pm i \cos 2\delta({\bi k})  & for $n^{\prime} = n + m + 1\; ({\rm mod}\; 2)$.}
\label{M}
\end{equation}
The contribution to the static spin susceptibility from inter-band scattering
that corresponds to true ($m=0$) 
or to hidden ($m=1$) N\'eel order
is then given by the Lindhard function
\begin{equation}
\chi_{\rm inter}(m,{\bi Q}_{\rm AF}) = -{1\over{a^2 N_{\rm Fe}}} \sum_{\bi k}
{n_F[\varepsilon_-({\bar{\bi k}})] - n_F[\varepsilon_+({\bi k})]\over
{\varepsilon_-({\bar{\bi k}})-\varepsilon_+({\bi k})}}
|{\cal M}_{+,{\bi k};-,{\bar{\bi k}}}|^2 ,
\label{Lndhrd}
\end{equation}
where $n_{\rm F}$ is the Fermi-Dirac distribution.  

\begin{figure}
\includegraphics{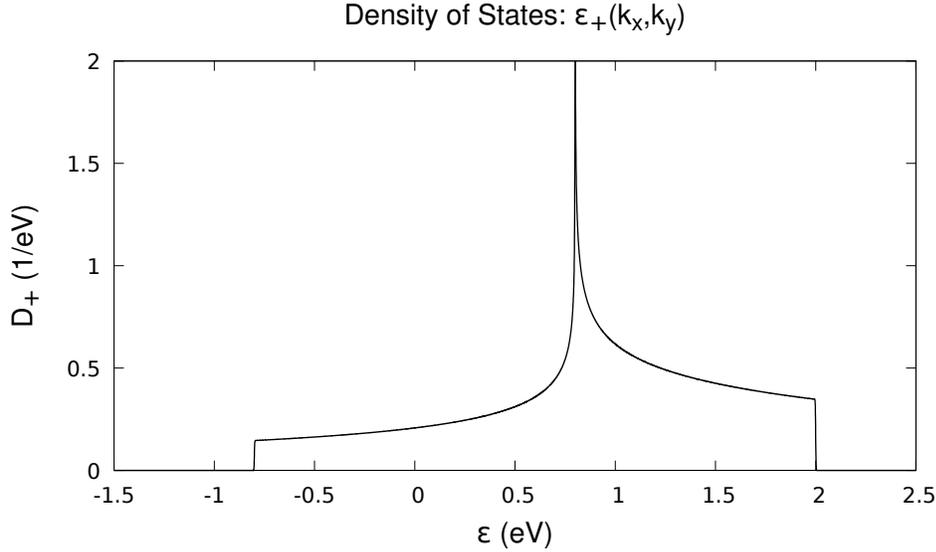}
\caption{Density of states of the bonding band evaluated numerically
at hopping parameters that are listed in the caption to Fig. \ref{FS0}:
$a^2 D_+(\varepsilon) = N_{\rm Fe}^{-1}\sum_{\bi k} \delta[\varepsilon-\varepsilon_+({\bi k})]$.
The one-iron Brillouin zone is divided into a $10,000\times 10,000$ grid,
while the $\delta$-function is approximated by
$(4 k_B T_0)^{-1} {\rm sech}^2 (\varepsilon/2 k_B T_0)$.
Here, $k_B T_0$ is $3$ parts in $10,000$ of the bandwidth.
The peak is a logarithmic van Hove singularity at $\varepsilon_+(\pi/a,\pi/a)$.}
\label{DoS}
\end{figure}

Next,
application of the perfect-nesting condition (\ref{prfct_nstng}) 
yields a more compact expression
for the inter-band contribution to the static spin susceptibility (\ref{Lndhrd}):
\begin{equation}
\chi_{\rm inter}(m,{\bi Q}_{\rm AF}) = {1\over{a^2 N_{\rm Fe}}} \sum_{\bi k}
{{1\over 2} - n_F[\varepsilon_+({\bi k})]
\over
{\varepsilon_+({\bi k})}}
|{\cal M}_{+,{\bi k};-,{\bar{\bi k}}}|^2.
\label{prfct_Lndhrd}
\end{equation}
We conclude that the static susceptibilities for true and for hidden N\'eel order diverge
logarithmically as 
$\chi_{\rm inter}(0,{\bi Q}_{\rm AF}) = {\rm lim}_{\epsilon\rightarrow 0}\, c^2D_+(0)\, {\rm ln}(W_{\rm bottom}/\epsilon)$
and
$\chi_{\rm inter}(1,{\bi Q}_{\rm AF}) = {\rm lim}_{\epsilon\rightarrow 0}\, s^2D_+(0)\, {\rm ln}(W_{\rm bottom}/\epsilon)$,
with corresponding density of states weighted by the 
magnitude square of the matrix element (\ref{M}):
\begin{eqnarray}
\label{ccD}
c^2D_+(\varepsilon) &=& (2\pi)^{-2} \int_{\rm BZ}d^2 k\, [\cos 2\delta({\bi k})]^2\delta[\varepsilon-\varepsilon_+({\bi k})] , \\
\label{ssD}
s^2D_+(\varepsilon) &=& (2\pi)^{-2} \int_{\rm BZ}d^2 k\, [\sin 2\delta({\bi k})]^2\delta[\varepsilon-\varepsilon_+({\bi k})] .
\end{eqnarray}
Above, $W_{\rm bottom} = - \varepsilon_+(0,0)$.
The weighted densities of states (\ref{ccD}) and (\ref{ssD}) are of comparable strength at the Fermi level,
 $\varepsilon = 0$.
For example, numerical calculations that are described in the caption to Fig. \ref{DoS}
yield the values
$c^2D_+(0) = 0.135\, a^{-2} {\rm eV}^{-1}$ and $s^2D_+(0) = 0.072\, a^{-2} {\rm eV}^{-1}$ 
for these quantities.
Here, hopping matrix elements coincide with those listed in the caption to Fig. \ref{FS0}.
Hidden magnetic order is therefore possible.

\section{Hidden magnetic order and excitations}
We have just seen how the perfect nesting of the Fermi surfaces 
shown by Fig. \ref{FS0} results in an instability towards long-range N\'eel order
per $d+$ and $d-$ orbital.  Below, we shall introduce an extended Hubbard model
over the square lattice that includes these orbitals alone.
Within mean field theory,
we shall see  that long-range hidden N\'eel order exists at half filling
because  of magnetic frustration by super-exchange interactions\cite{jpr_ehr_09,jpr_10}.

\subsection{Extended Hubbard model}
We shall now add on-site interactions due to Coulomb repulsion
and super-exchange interactions via the Se atoms to
the hopping Hamiltonian (\ref{hop}).
The Hamiltonian then has three parts:
$H = H_{\rm hop} + H_U + H_{\rm sprx}$.
On-site Coulomb repulsion is counted by the second term\cite{2orb_Hbbrd},
\begin{eqnarray}
H_U =  \sum_i &[U_0 n_{i,\alpha,\uparrow} n_{i,\alpha,\downarrow}
                +J_0 {\bi S}_{i, d-}\cdot {\bi S}_{i, d+} \nonumber \\
                &+U_0^{\prime} n_{i,d+} n_{i,d-}
                +J_0^{\prime} (c_{i,d+,\uparrow}^{\dagger}c_{i,d+,\downarrow}^{\dagger}
                            c_{i,d-,\downarrow}c_{i,d-,\uparrow}+ {\rm h.c.})],
\label{U}
\end{eqnarray}
where $n_{i,\alpha,s} = c_{i,\alpha,s}^{\dagger}c_{i,\alpha,s}$ is the occupation
operator, 
where ${\bi S}_{i,\alpha} = \sum_{s,s^{\prime}}
c_{i,\alpha,s}^{\dagger}{\boldsymbol \sigma}_{s,s^{\prime}} c_{i,\alpha,s^{\prime}}$
is the spin operator,
and where $n_{i,\alpha} = n_{i,\alpha,\uparrow} + n_{i,\alpha,\downarrow}$.
Above, $U_0>0$ denotes the intra-orbital on-site Coulomb repulsion energy,
while $U_0^{\prime} > 0$ denotes the inter-orbital one.
Also, $J_0 < 0$ is the Hund's Rule exchange coupling constant, which is ferromagnetic,
while $J_0^{\prime}$ denotes the matrix element for on-site-orbital Josephson tunneling.

The third and last term in the Hamiltonian represents super-exchange interactions
among the iron spins via the selenium atoms:
\begin{eqnarray}
H_{\rm sprx} =
\sum_{\langle i,j \rangle} & J_1^{({\rm sprx})} ({\bi S}_{i, d-} + {\bi S}_{i, d+})
\cdot ({\bi S}_{j, d-} + {\bi S}_{j, d+}) \nonumber \\
&+\sum_{\langle\langle  i,j \rangle\rangle} J_2^{({\rm sprx})} ({\bi S}_{i, d-} + {\bi S}_{i, d+})
\cdot ({\bi S}_{j, d-} + {\bi S}_{j, d+}).
\label{sprx}
\end{eqnarray}
Above, $J_1^{({\rm sprx})}$ and $J_2^{({\rm sprx})}$ are positive 
super-exchange coupling constants
over nearest neighbor and next-nearest neighbor iron sites.
We shall assume henceforth that magnetic frustration is moderate to strong:
$J_2^{({\rm sprx})} > 0.5 J_1^{({\rm sprx})}$.
In isolation, the  above term of the Hamiltonian then
favors ``stripe'' spin-density wave order at half filling over conventional N\'eel order.

\subsection{Mean field theory}\label{mft}
Following the mean-field treatment of antiferromagnetism in
the conventional Hubbard model over the square lattice
at half filling\cite{hirsch_85,schrieffer_wen_zhang_89,singh_tesanovic_90,chubokov_frenkel_92}, 
assume that the expectation value of the magnetic moment per site, per orbital,
shows hidden N\'eel order: 
\begin{equation}
\langle m_{i,\alpha}\rangle = (-1)^{\alpha} e^{i{\bi Q}_{\rm AF}\cdot{\bi r}_i} \langle m_{0,0}\rangle,
\label{ordered_moment}
\end{equation}
where
$\langle m_{i,\alpha}\rangle = {1\over 2}
\langle n_{i,\alpha,\uparrow}\rangle-{1\over 2}\langle n_{i,\alpha,\downarrow}\rangle$.
Previous calculations in the local-moment limit (\ref{hsnbrg}) indicate that
the above hidden magnetic order is more stable than the ``stripe'' spin-density wave (SDW) mentioned above
at weak to moderate strength in the Hund's Rule coupling\cite{jpr_17,jpr_10}.
The super-exchange terms, $H_{\rm sprx}$, 
make no contribution within the mean-field approximation,
since the net magnetic moment per iron atom is null in the hidden-order N\'eel state.  
And we shall neglect the on-site Josephson tunneling term in (\ref{U}) $H_U$.
This is valid at the strong-coupling limit, $U_0\rightarrow\infty$,
where the formation of a spin singlet per iron-site-orbital is suppressed.
We are then left
with the two on-iron--site repulsion terms and the Hund's Rule term in $H_U$.

The mean-field replacement of the intra-orbital on-site term ($U_0$)
is the standard one\cite{hirsch_85}.  In particular, replace
\begin{eqnarray}
n_{i,\alpha,\uparrow} n_{i,\alpha,\downarrow} \rightarrow 
{1\over 2}\langle n_{i,\alpha}\rangle (n_{i,\alpha,\uparrow}+n_{i,\alpha,\downarrow})
&-\langle m_{i,\alpha}\rangle (n_{i,\alpha,\uparrow} - n_{i,\alpha,\downarrow})\nonumber\\
&-\langle n_{i,\alpha,\uparrow}\rangle \langle n_{i,\alpha,\downarrow}\rangle.\nonumber
\end{eqnarray}
The first term above can be absorbed into the chemical potential
because $\langle n_{i,d-}\rangle = \langle n_{i,d+}\rangle$,
the last term above is a constant energy shift, leaving a mean-field contribution
to the Hamiltonian $-\sum_i\sum_{\alpha}  U_0\langle m_{i,\alpha}\rangle (n_{i,\alpha,\uparrow} - n_{i,\alpha,\downarrow})$.
The mean-field replacement of the inter-orbital on-iron-site repulsion term
($U_0^{\prime}$) in $H_U$
is also standard:
$$n_{i,d+} n_{i,d-} \rightarrow  n_{i,d+} \langle n_{i,d-}\rangle
+ \langle n_{i,d+}\rangle n_{i,d-} - \langle n_{i,d+}\rangle \langle n_{i,d-}\rangle.$$
The first two terms above can again be absorbed into a shift of the chemical potential,
while the third and last term above is again a constant energy shift.  
The inter-orbital repulsion term, hence,
makes no contribution to the Hamiltonian within the mean-field approximation.
Last, we make the same type of mean-field replacement for the Hund's Rule
term ($J_0$) in $H_U$:
$${\bi S}_{i,d+} \cdot {\bi S}_{i,d-} \rightarrow 
 S_{i,d+}^{(z)} \langle S_{i,d-}^{(z)}\rangle 
+ \langle S_{i,d+}^{(z)}\rangle S_{i,d-}^{(z)}
- \langle S_{i,d+}^{(z)}\rangle \langle S_{i,d-}^{(z)}\rangle .$$
Again, the last term above is just a constant energy shift.
The first two terms, however, contribute to the mean-field Hamiltonian:
$\sum_i \sum_{\alpha} {1\over 2} J_0 \langle m_{i,{\bar\alpha}}\rangle (n_{i,\alpha,\uparrow} - n_{i,\alpha,\downarrow})$,
which is equal to 
$-\sum_i \sum_{\alpha} {1\over 2} J_0 \langle m_{i,\alpha}\rangle (n_{i,\alpha,\uparrow} - n_{i,\alpha,\downarrow})$
in the case of hidden magnetic order (\ref{ordered_moment}).
Here, $\overline {d\pm} = d\mp$.

The net contribution to the mean-field Hamiltonian from interactions
in the present two-orbital Hubbard model is then
$$-\sum_i \sum_{\alpha} U(\pi) \langle m_{i,\alpha}\rangle (n_{i,\alpha,\uparrow} - n_{i,\alpha,\downarrow})
= - \langle m_{0,0} \rangle  U(\pi)
\sum_i \sum_{\alpha}
(-1)^{\alpha} e^{i{\bi Q}_{\rm AF}\cdot{\bi r}_i} 
(n_{i,\alpha,\uparrow} - n_{i,\alpha,\downarrow}),$$
where 
\begin{equation}
U(\pi) = U_0+{1\over 2} J_0.
\label{U(pi)}
\end{equation}
Notice that the last sum above is simply twice the hidden ($m=1$) ordered moment
$S_z (m,{\bi Q}_{\rm AF})$ defined by (\ref{Sz}).
Inspection of (\ref{sz}) then yields that the mean-field Hamiltonian for
the present two-orbital Hubbard model takes the form
\begin{eqnarray}
H^{(mf)} = & \sum_s\sum_{\bi k}\sum_{n} \varepsilon_n({\bi k}) c_s^{\dagger}(n,{\bi k}) c_s(n,{\bi k})  \nonumber \\
& \mp \sum_s\sum_{\bi k}[({\rm sgn}\, s) \Delta({\bi k}) 
c_s^{\dagger}(1,{\bar{\bi k}}) c_s(2,{\bi k})+{\rm h.c.}],
\label{Hmf}
\end{eqnarray}
with a gap function
\begin{equation}
\Delta({\bi k}) = \Delta_0 \sin[2\delta({\bi k})],
\label{gap}
\end{equation}
where ${\bar{\bi k}} = {\bi k}+{\bi Q}_{\rm AF}$,
and where
\begin{equation}
\Delta_0 = \langle m_{0,0}\rangle U(\pi).
\label{Delta0}
\end{equation}
Here, we have used the result (\ref{M}) for the matrix element
in the case of hidden magnetic order ($m=1$).  
Here also,
intra-band scattering
($n^{\prime} = n$) has been neglected because it shows no nesting.
After shifting the sum in momentum of the first term in (\ref{Hmf})
by ${\bi Q}_{AF}$ for the anti-bonding band, $n=1$, we arrive at the
final form of the mean-field Hamiltonian:
\begin{eqnarray}
H^{(mf)} = & \sum_s\sum_{\bi k}\varepsilon_+({\bi k}) [c_s^{\dagger}(2,{\bi k}) c_s(2,{\bi k})
- c_s^{\dagger}(1,{\bar{\bi k}}) c_s(1,{\bar{\bi k}})]  \nonumber \\
&  +  \sum_s\sum_{\bi k}[({\rm sgn}\, s)\Delta({\bi k}) 
c_s^{\dagger}(1,{\bar{\bi k}}) c_s(2,{\bi k})+{\rm h.c.}].
\label{HMF}
\end{eqnarray}
Above, we have set the $\pm$ sign in the matrix element (\ref{M})
to minus for convenience.

The mean-field Hamiltonian (\ref{HMF}) is diagonalized 
in the standard way by writing the electron in terms of new quasi-particle 
excitations\cite{schrieffer_wen_zhang_89,singh_tesanovic_90,chubokov_frenkel_92}:
%
\begin{eqnarray}
c_s^{\dagger}(2,{\bi k}) &=& u({\bi k}) \alpha_s^{\dagger}(2,{\bi k})
 - ({\rm sgn}\, s) v({\bi k}) \alpha_s^{\dagger}(1,{\bar{\bi k}}) , \nonumber\\
c_s^{\dagger}(1,{\bar{\bi k}}) &=& ({\rm sgn}\, s) v({\bi k})  \alpha_s^{\dagger}(2,{\bi k}) 
 + u({\bi k}) \alpha_s^{\dagger}(1,{\bar{\bi k}}) .
\label{qps}
\end{eqnarray}
%
Above, $u({\bi k})$ and $v({\bi k})$ are coherence factors with square magnitudes
\begin{equation}
u^2 = {1\over 2} + {1\over 2} {\varepsilon_+\over E} \quad{\rm and}\quad
v^2 = {1\over 2} - {1\over 2} {\varepsilon_+\over E},
\label{u&v}
\end{equation}
where $E({\bi k}) = [\varepsilon_+^2({\bi k}) + \Delta^2({\bi k})]^{1/2}$.
The mean-field Hamiltonian can then be expressed in terms of the occupation
of quasiparticles by
\begin{equation}
H^{(mf)} = \sum_s\sum_{\bi k} E({\bi k})[\alpha_s^{\dagger}(2,{\bi k}) \alpha_s(2,{\bi k}) 
- \alpha_s^{\dagger}(1,{\bar{\bi k}}) \alpha_s (1,{\bar{\bi k}})].
\label{dgnl}
\end{equation}
The quasi-particle excitation energies are then $E({\bi k})$ for particles
and $E({\bar {\bi k}})$ for holes.
Notice that the gap (\ref{gap}) in the excitation spectrum has $D_{xy}$ symmetry.
(See Fig. \ref{FS0}.)
At half filling then, the energy band $-E({\bar {\bi k}})$ is filled,
while the energy band $+E({\bi k})$ is empty.
Last, inverting (\ref{qps}) yields
%
\begin{eqnarray}
\alpha_s^{\dagger}(2,{\bi k}) &=& u({\bi k}) c_s^{\dagger}(2,{\bi k})
 + ({\rm sgn}\, s) v({\bi k}) c_s^{\dagger}(1,{\bar{\bi k}}) , \nonumber\\
\alpha_s^{\dagger}(1,{\bar{\bi k}}) &=& - ({\rm sgn}\, s)v({\bi k})  c_s^{\dagger}(2,{\bi k}) 
 + u({\bi k}) c_s^{\dagger}(1,{\bar{\bi k}}) .
\label{QPS}
\end{eqnarray}
%
As expected, the quasiparticles are a coherent superposition of an electron
of momentum ${\bi k}$ in the bonding band $2$
with an electron of momentum ${\bi k}+{\bi Q}_{\rm AF}$ 
in the anti-bonding band $1$.

\begin{figure}
\includegraphics{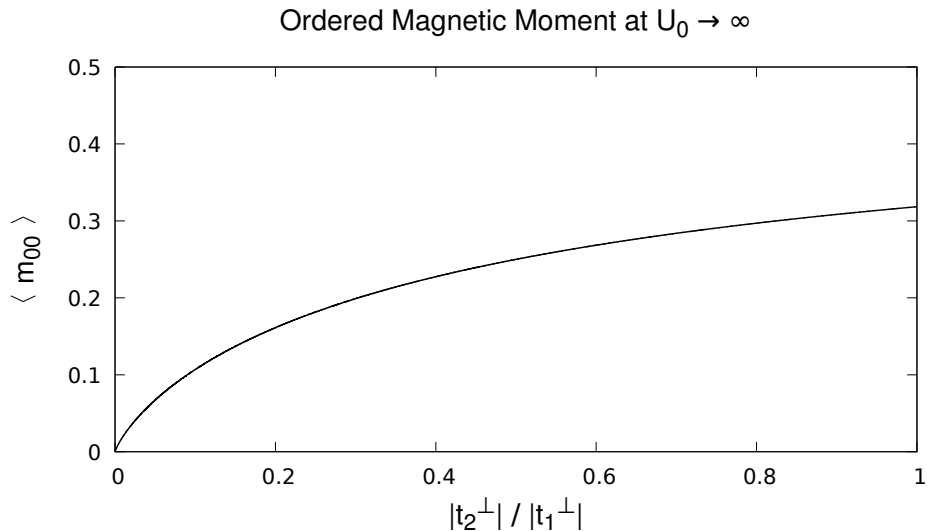}
\caption{Shown is the ordered magnetic moment at $U_0\rightarrow\infty$:
$\langle m_{0,0}\rangle = (2 N_{\rm Fe})^{-1} \sum_{\bi k} |\sin\,2\delta({\bi k})|$.
The one-iron Brillouin zone is divided into a $10,000\times 10,000$ grid.}
\label{magnetic_moment}
\end{figure}

Finally, to obtain the gap equation, 
we exploit the pattern of hidden N\'eel order
(\ref{ordered_moment}), 
and equivalently write the gap maximum (\ref{Delta0}) as
$$\Delta_0 = 
{\cal N}^{-1}\sum_i\sum_{\alpha} U(\pi) \langle m_{i,\alpha}\rangle (-1)^{\alpha} e^{i{\bi Q}_{\rm AF}\cdot {\bi r}_i} 
= {\cal N}^{-1} U(\pi) \langle S_z(1,{\bi Q}_{\rm AF})\rangle.$$
Using expression (\ref{sz}) and the result (\ref{M}) 
in the case of hidden magnetic order ($m=1$) yields the relationship
$$\Delta_0 = -
{\cal N}^{-1}{1\over 2}\sum_s\sum_{\bi k}\sum_{n} U(\pi) ({\rm sgn}\, s) [\sin\, 2\delta({\bi k})]
\langle c_s^{\dagger}({\bar n},{\bar{\bi k}}) c_s(n,{\bi k})\rangle,$$
where ${\bar n} = 1 + (n\; {\rm mod}\; 2)$.
Again, we have neglected intra-band scattering.
Also,
notice that the sums over the bands (\ref{plane_waves}) and the spins yield
the hSDW order parameter
\begin{equation}
i \langle c_s^{\dagger}(d_{y({\bar\delta})z},{\bar{\bi k}}) c_s(d_{x(\delta)z},{\bi k}) 
-  c_s^{\dagger}(d_{x({\bar\delta})z},{\bar{\bi k}}) c_s(d_{y(\delta)z},{\bi k}) \rangle ({\rm sgn}\, s),
\end{equation}
\label{OP}
which is orbitally isotropic.
Substituting in (\ref{qps}) and the conjugate annihilation operators,
and recalling that the $n=1$ quasi-particle band is filled in the groundstate, 
while the $n=2$ quasi-particle band is empty,
yields 
$\langle c_s^{\dagger}({\bar n},{\bar{\bi k}}) c_s(n,{\bi k})\rangle
 = - ({\rm sgn}\, s) u({\bi k}) v({\bi k})$
for the expectation value.
We thereby obtain the relationship
$$\Delta_0 = 
{\cal N}^{-1}\sum_{\bi k} U(\pi)
[\sin\, 2\delta({\bi k})] \Delta({\bi k})/E({\bi k}),$$
or equivalently, the gap equation
\begin{equation}
1 = U(\pi) {\cal N}^{-1} \sum_{\bi k} 
{[\sin\, 2\delta({\bi k})]^2 \over{\sqrt{\varepsilon_+^2({\bi k})+\Delta_0^2 [\sin\, 2\delta({\bi k})]^2}}}.
\label{gap_eq}
\end{equation}
In the limit $\Delta_0\rightarrow\infty$, we then have
$\Delta_0 = U(\pi) {\cal N}^{-1} \sum_{\bi k} |\sin\, 2\delta({\bi k})|$,
which yields a hidden-order moment
$\langle m_{0,0}\rangle = {\cal N}^{-1} \sum_{\bi k} |\sin\, 2\delta({\bi k})|$
bounded by $1/2$.
In the special case
$|t_1^{\perp}| = 2 |t_2^{\perp}|$,
inspection of (\ref{s_2dlt}) yields
$|\sin\, 2\delta({\bi k})| = |\sin\, k_x a| |\sin\, k_y a|/[1-(\cos\, k_x a)(\cos\, k_y a)]$.
In the thermodynamic limit, $N_{\rm Fe}\rightarrow\infty$,
integration along one of the principal axes followed by a series expansion in turn yields 
$\langle m_{0,0}\rangle = 1/4$
for the previous expression.
In the limit $|t_1^{\perp}|\rightarrow 0$, on the other hand,
we have $|\sin\, 2\delta({\bi k})| \rightarrow 1$ by (\ref{s_2dlt}),
which yields an ordered magnetic moment
$\langle m_{0,0}\rangle = 1/2$.
Figure \ref{magnetic_moment} shows the ordered magnetic moment at $U_0\rightarrow\infty$
versus hybridization of the $3d_{xz}$ and $3d_{yz}$ orbitals.

The above mean field theory predicts quasiparticles with excitation energies
that disperse as $E_+({\bi k})$ and $E_-({\bi k})$,
where
$E_{\pm}({\bi k}) =
(\varepsilon_{\pm}^2({\bi k})+\Delta_0^2 [\sin 2\delta({\bi k})]^2)^{1/2}$.
They reach zero at point nodes located where the Fermi surface crosses
a principal axis, at which the phase shift $\delta({\bi k})$
is a multiple of $\pi/2$. These point nodes are shown by Fig. \ref{FS0}.
The quasi-particle energy spectra disperse about the nodes
in a Dirac-cone fashion:
$E({\bi k}) \cong 
[v_F^2 k_{\perp}^2 + (2 \Delta_0 \delta_D^{\prime})^2 k_{\parallel}^2]^{1/2}$,
where $v_F$ is the Fermi velocity of $\varepsilon_{\pm}({\bi k})$
at the node, and where $\delta_D^{\prime}$ is the gradient of the phase shift
$\delta({\bi k})$ at the node.
Here $(k_{\parallel},k_{\perp})$ denote the momentum coordinates about a 
point node in the directions parallel and perpendicular to the Fermi surface there.
Last, notice that combining the spectra $E_+({\bi k})$ and $E_-({\bi k})$
in the folded Brillouin zone results in four Dirac cones at the Fermi level.
(C.f. ref. \cite{ran_09}).

\subsection{Low-energy collective modes}\label{cllctv_mds}\label{hydro}
The groundstate of the above mean field theory is the filled 
energy band $-E({\bar{\bi k}})$:  
$|\Psi_0\rangle = \prod_{{\bi k},s} \alpha_s^{\dagger}(1,{\bi k}) |0\rangle$.
Inspection of (\ref{QPS}) yields that it can also be expressed as
\begin{equation}
|\Psi_0\rangle = \prod_{{\bi k},s} 
[u({\bar{\bi k}}) - ({\rm sgn}\, s)v({\bar{\bi k}})  c_s^{\dagger}(2,{\bar{\bi k}}) c_s(1,{\bi k})]
|1\rangle ,
\label{Psi_0}
\end{equation}
where $|1\rangle = \prod_{{\bi k},s} c_s^{\dagger}(1,{\bi k})|0\rangle$
is the filled anti-bonding band of non-interacting electrons.
Next, observe that each pair of factors above per momentum
over  spin $\uparrow$ and $\downarrow$ can be expressed as
$u^2({\bar{\bi k}})\,{\rm exp}(-[v({\bar{\bi k}})/u({\bar{\bi k}})]{\hat{\bi n}}\cdot \sum_{s_1,s_2} 
c_{s_1}^{\dagger}(2,{\bar{\bi k}}) {\boldsymbol \sigma}_{s_1,s_2}  c_{s_2}(1,{\bi k}))$,
with unit vector ${\hat{\bi n}}$ along with the $z$ axis.
Now define a new spin quantization axis $z^{\prime} = y$, 
along with the remaining axes $x^{\prime} = z$ and $y^{\prime} = x$.
If, more generally, we let the axis of the sub-lattice magnetization
${\hat{\bi n}}$ lie in the $z$-$x$ plane, then the spin operator
in the argument of the exponential above becomes
$$\sum_{s_1,s_2}
c_{s_1}^{\prime\, \dagger}(2,{\bar{\bi k}}) 
[ (\cos \,\phi) \sigma_x + (\sin\, \phi) \sigma_y]_{s_1,s_2}
c_{s_2}^{\prime}(1,{\bi k}).$$ 
Here, $c_s^{\prime}$ and $c_s^{\prime\, \dagger}$
are the electron annihilation and creation operators
in the new quantization axes.  Here also, $\phi$ is the angle that ${\hat{\bi n}}$
makes with the $z$ axis. Re-expanding the exponential operator above then yields
the groundstate (\ref{Psi_0}) in the new quantization axes:
\begin{eqnarray}
|\Psi_0 (\phi) \rangle
= \prod_{{\bi k}}
& [u({\bar{\bi k}}) - e^{-i\phi} v({\bar{\bi k}})  c_{\uparrow}^{\prime\, \dagger}(2,{\bar{\bi k}}) c_{\downarrow}^{\prime}(1,{\bi k})] \nonumber\\
&\cdot[u({\bar{\bi k}}) - e^{+i\phi} v({\bar{\bi k}})  c_{\downarrow}^{\prime\, \dagger}(2,{\bar{\bi k}}) c_{\uparrow}^{\prime}(1,{\bi k})]
|1\rangle .
\label{GS}
\end{eqnarray}
It has indefinite spin $S_y$ along the new quantization axis:
\begin{equation}
|\Psi_0 (\phi) \rangle 
= \sum_{m=-N_{\rm Fe}}^{N_{\rm Fe}} e^{-i m \phi} |\Psi_0^{(m)}\rangle,
\label{gs}
\end{equation}
where $|\Psi_0^{(m)}\rangle$ are projections of the groundstate 
that have spin $S_y$ equal to $m\hbar$.
Then because
$S_y = i \hbar {\partial\over{\partial\phi}}$,
we have that the  macroscopic phase angle and the macroscopic spin along
the $y$ axis satisfy the commutation
relationship\cite{anderson_64} $[\phi, S_y] = -i\hbar$.

Define, next, the macroscopic magnetization, $M_y = S_y / V$, where $V$ is the area,
and let it and the phase angle $\phi$ vary slowly over the bulk.
Their dynamics is then governed by 
the hydrodynamic Hamiltonian\cite{halperin_hohenberg_69,forster_75}
$H_{\rm hydro}  = \int d^2 r\, {\cal H}_{\rm hydro}$, with the Hamiltonian density
\begin{equation}
{\cal H}_{\rm hydro} = 
{1\over{2\chi_{\perp}}} M_y^2 + {1\over 2} \rho_s |{\bi\nabla} \phi|^2.
\label{H_hydro}
\end{equation}
Above, $\chi_{\perp}$ and $\rho_s$ denote, respectively, 
the transverse spin susceptibility and the spin stiffness of
the present hidden spin-density wave state.
Given the commutation relationship
$[\phi({\bi r}), M_y({\bi r}^{\prime})] = 
-i\hbar\, \delta^{(2)}({\bi r}-{\bi r}^{\prime})$,
we obtain the following dynamical equations:
\begin{eqnarray}
{\dot M_y} &=& - \rho_s \nabla^2 \phi ,
\nonumber \\
{\dot \phi} &=& - M_y / \chi_{\perp}.
\label{dynamics}
\end{eqnarray}
The magnetization thus satisfies the wave equation
${\ddot M_y} = c_0^2 \nabla^2 M_y$, with propagation velocity
$c_0 = (\rho_s/\chi_{\perp})^{1/2}$.  We conclude that the
present hidden spin-density wave state supports antiferromagnetic
spin-wave excitations  that disperse acoustically in frequency:
$\omega({\bar{\bi k}}) = c_0 |{\bi k}|$. And since the above
dynamics can be rotated by $90$ degrees about the $z$ axis, 
there then exist {\it two} acoustic spin-wave excitations per momentum.

\subsection{Transverse spin susceptibility and spin rigidity}\label{rho_chi}
To compute the transverse spin susceptibility, 
we apply an external magnetic field along the $y$ axis by adding the term
$-h \sum_i \sum_{\alpha} S_{i,\alpha}^{(y)}$ to the Hamiltonian
$H_{\rm hop} + H_U + H_{\rm sprx}$.
The on-site-orbital repulsion terms, the Hund's Rule coupling terms,
and the super-exchange terms can then be replaced by the isotropic
mean-field approximations
\begin{eqnarray}
n_{i,\alpha,\uparrow} n_{i,\alpha,\downarrow} \rightarrow 
{1\over 2}\langle n_{i,\alpha}\rangle (n_{i,\alpha,\uparrow}+n_{i,\alpha,\downarrow})
&-2 \langle {\bi m}_{i,\alpha}\rangle \cdot {\bi S}_{i,\alpha}\nonumber\\
&-\langle n_{i,\alpha,\uparrow}\rangle \langle n_{i,\alpha,\downarrow}\rangle\nonumber
\end{eqnarray}
and
$${\bi S}_{i,\alpha} \cdot {\bi S}_{j,\beta} \rightarrow 
 {\bi S}_{i,\alpha} \cdot \langle {\bi S}_{j,\beta}\rangle 
+ \langle {\bi S}_{i,\alpha}\rangle \cdot {\bi S}_{j,\beta}
- \langle {\bi S}_{i,\alpha}\rangle \cdot \langle {\bi S}_{j,\beta}\rangle .$$
Yet the external magnetic field cants the antiferromagnetically aligned moments
per orbital along the $y$ axis by the transverse magnetization per orbital,
$\langle m_{\perp}\rangle$.  It makes a contribution to the above mean-field
replacements that can be accounted for by making the replacement
$h\rightarrow h + 2 U(0) \langle m_{\perp}\rangle$ in 
the paramagnetic term that we added above to the Hamiltonian, where
\begin{equation}
U(0) = U_0 - {1\over 2}J_0 - 4 J_1^{({\rm sprx})} - 4 J_2^{({\rm sprx})}.
\label{U(0)}
\end{equation}
We thereby arrive at the formula
\begin{equation}
\chi_{\perp} = {\chi_{\perp}^{(0)}\over{1 - a^2 U(0) \chi_{\perp}^{(0)}}}
\label{RPA}
\end{equation}
for the transverse spin susceptibility per iron atom,
where $\chi_{\perp}^{(0)}$ is the naive transverse spin susceptibility
that neglects the effect of canting.

The formula for the naive transverse spin susceptibility 
is well known\cite{denteneer_an_van_leeuwen_93}, and it is derived in
\ref{ppndx_b}.  It reads
\begin{equation}
\chi_{\perp}^{(0)} = {1\over 2} (a^2 N_{\rm Fe})^{-1} \sum_{\bi k} 
{\Delta^2({\bi k})\over{[\varepsilon_+^2({\bi k})+\Delta^2({\bi k})]^{3/2}}}.
\label{chi_perp_0}
\end{equation}
At the weak-coupling limit,
$U(0), U(\pi) \rightarrow 0$,
the transverse spin susceptibility (\ref{RPA}) is given by $\chi_{\perp}^{(0)}$,
and the quotient in the sum over momentum above is equal to 
$2\,\delta[\varepsilon_+({\bi k})]$.  We thereby obtain the Pauli paramagnetic
susceptibility at weak-coupling,
\begin{equation}
{\rm lim}_{U(0),U(\pi)\rightarrow 0} \chi_{\perp} = D_+(0),
\label{wk_cplng}
\end{equation}
where $D_+(\varepsilon)$ is the density of states of the bonding band
$\varepsilon_+({\bi k})$.  (See Fig. \ref{DoS}.)

At the strong-coupling limit, $U_0 \rightarrow \infty$,
it's useful to re-write the formula (\ref{RPA}) as\cite{denteneer_an_van_leeuwen_93}
\begin{equation}
\chi_{\perp} = {1\over{a^2 U(0)}} \Biggl[{1\over{1 - a^2 U(0) \chi_{\perp}^{(0)}}}-1\Biggr].
\label{rpa}
\end{equation}
Observe, next, the following identity for the quotient in (\ref{chi_perp_0}):
$${\Delta^2({\bi k})\over{E^3({\bi k})}} = 
{[\sin\, 2 \delta({\bi k})]^2\over{E({\bi k})}}+
{[\cos\, 2 \delta({\bi k})]^2\over{E({\bi k})}}
-{\varepsilon_+^2(\bi k)\over{E^3({\bi k})}}.$$
Applying the gap equation (\ref{gap_eq}) then yields the result
$a^2 U(\pi) \chi_{\perp}^{(0)} = 1 + I_1 - I_2$, where
\numparts
\begin{eqnarray}
\label{intrmdt_a}
I_1 &=& {1\over 2} N_{\rm Fe}^{-1}\sum_{\bi k}
 {[\cos 2\delta({\bi k})]^2\over{E({\bi k})}} U(\pi),\\
\label{intrmdt_b}
I_2 &=& {1\over 2} N_{\rm Fe}^{-1}\sum_{\bi k} 
 {\varepsilon_+^2{\bi k})\over{E^3({\bi k})}} U(\pi).
\end{eqnarray}
\endnumparts
In the limit $U_0\rightarrow\infty$,
it can be shown that $I_1 = I_2$ 
as the pure-imaginary hopping matrix element $t_{2}^{\perp}$ tends to zero.
(See \ref{ppndx_b}.)  In such case, $a^2 U(\pi) \chi_{\perp}^{(0)} = 1$, 
and we thereby achieve the result
\begin{eqnarray}
{\rm lim}_{U_0\rightarrow\infty} \chi_{\perp} & = & 
{1\over{a^2 U(0)}} \Biggl(\Biggl[1-{U(0)\over{U(\pi)}}\Biggr]^{-1}-1\Biggr) \nonumber\\
& = & a^{-2} [J_0 + 4 J_1^{({\rm sprx})} + 4 J_2^{({\rm sprx})}]^{-1}.
\label{strng_cplng}
\end{eqnarray}
It coincides precisely with the corresponding result 
for the Heisenberg model (\ref{hsnbrg}) \cite{jpr_10}
after making the assignments
$J_1^{\parallel} = J_1^{({\rm sprx})}$ and
$J_2^{\perp} = J_2^{({\rm sprx})}$
for two of the exchange coupling constants.
Here $J_n^{\parallel}$ and $J_n^{\perp}$ represent intra-orbital and inter-orbital
Heisenberg exchange coupling constants among the $d+$ and $d-$ orbitals.

Next, to compute the spin stiffness $\rho_s$ at half filling,
we follow the calculation of the same
quantity in the case of the conventional Hubbard model over
the square lattice\cite{schrieffer_wen_zhang_89,singh_tesanovic_90,chubokov_frenkel_92,denteneer_an_van_leeuwen_93,shi_singh_95}.
At zero temperature, the spin rigidity saturates the f-sum rule for the spin current.
We therefore arrive at the expression
\begin{equation}
\rho_s =  {1\over{N_{\rm Fe}}} \sum_{\bi k} [u^2({\bi k})-v^2({\bi k})] {1\over 4}\varepsilon_{\parallel}({\bi k})
\label{rho_s}
\end{equation}
for it.  Here, we have taken the average over the two principal axes.
In the weak-coupling limit, where the gap function vanishes,
we therefore get
${\rm lim}_{U(\pi)\rightarrow 0}\, \rho_s 
= N_{\rm Fe}^{-1}\sum_{\bi k}^{\prime} 
t_1^{\parallel} (\cos \, k_x a + \cos \, k_y a)$,
where the prime notation indicates the condition that $\varepsilon_+({\bi k}) < 0$.
In such case, the sum over momenta lies inside the Fermi surface
at the center of the Brillouin zone. (See Fig. \ref{FS0}.)
At strong coupling $U(\pi)\rightarrow \infty$, on the other hand, it is useful
to return to the original expression (\ref{rho_s}):
\begin{equation}
\rho_s =  {1\over{N_{\rm Fe}}} \sum_{\bi k} {1\over 4}
{\varepsilon_+({\bi k})\cdot\varepsilon_{\parallel}({\bi k})
\over{\sqrt{\varepsilon_+^2({\bi k})+\Delta_0^2 [\sin \, 2 \delta({\bi k})]^2}}}.
\label{RHO_S}
\end{equation}
Here, we have substituted in the expressions for the coherence factors (\ref{u&v}).
Approximate now all dispersions in energy about the Dirac nodes at the
Fermi surface $\varepsilon_+({\bi k}) = 0$: e.g.;
$\varepsilon_+({\bi k}) \cong v_F (k_x - k_D)$,
$\varepsilon_{\parallel} ({\bi k}) \cong \varepsilon_{\parallel}(k_D,0)+ v_{\parallel} (k_x - k_D)$, and
$\sin\, 2 \delta({\bi k}) \cong 2 \delta_D^{\prime} k_y$,
where the coordinates of the Dirac node are $(k_D,0)$.
After taking the thermodynamic limit, $N_{\rm Fe}\rightarrow \infty$, and
after cutting off the resulting integrals in momentum 
by $k_1\sim k_D/2$ in both the $x$ and
in the $y$ directions, we obtain the following result in the limit of strong coupling:
\begin{equation}
{\rm lim}_{U(\pi)\rightarrow\infty}\, \rho_s  =
{2\over 3} {(k_1 a)^3\over{(2\pi)^2}} 
{v_{\rm F} v_{\parallel}\over{\delta_D^{\prime} \Delta_0}} {1\over a}
{\rm ln}\Bigl(2 e^{1/3} {2\delta_D^{\prime} \Delta_0 \over{v_F}}\Bigr).
\label{rho_s_strong}
\end{equation}
In this limit, the spin stiffness at half filling therefore scales as
$\rho_s\sim (t^2/U) {\rm ln}(U/t)$ with the scale of the hopping matrix elements
$t$, and with the scale of the on-site repulsion energy $U$.
It is useful to compare the latter result for the rigidity of hidden magnetic order
at strong on-site-orbital repulsion (\ref{rho_s_strong})
with that obtained from the corresponding two-orbital Heisenberg model (\ref{hsnbrg}) \cite{jpr_10}:
$\rho_s = 2 s_0^2 
[J_1^{\parallel}-J_1^{\perp}+ 2 (J_2^{\perp}-J_2^{\parallel})]$.
It yields the assignments
$J_1^{\perp} = J_1^{({\rm sprx})} -  \rho_s / 2 s_0^2$
and
$J_2^{\parallel} = J_2^{({\rm sprx})}$
for the remaining two exchange coupling constants.

\section{Eliashberg theory}
The previous mean-field approximation of the
extended two-orbital Hubbard model 
for a single layer of heavily electron-doped iron-selenide
predicts Dirac quasi-particle excitations 
at nodes where the Fermi surface crosses a principal axis.
(See Fig. \ref{FS0}.)
Below, we shall demonstrate how the Fermi surface at weak coupling
experiences a Lifshitz transition to Fermi-surface  pockets at
the corner of the two-iron Brillouin zone as the
on-site-orbital repulsion grows strong.
We will achieve this by first formulating an Eliashberg theory
for the extended Hubbard model in the electron-hole channel.

\subsection{Hidden spinwaves and interaction with electrons}
It was revealed in section \ref{cllctv_mds} that the above mean field theory
for the hidden N\'eel state of the Hubbard model over the square lattice
harbors spin-wave excitations that collapse to zero energy
at the N\'eel wavenumber ${\bi Q}_{\rm AF}$.  
The hidden N\'eel state of the corresponding Heisenberg model over the square lattice 
exhibits the very same  hidden spin-wave excitations\cite{jpr_10}.
Consider then the propagator for hidden spinwaves:
$iD({\bi q},\omega) = 
\langle {1\over{\sqrt{2}}} m^{+}(\pi) 
{1\over{\sqrt{2}}} m^{-}(\pi)\rangle |_{{\bi q},\omega}$,
where
$m^{\pm} (\pi) = m_x (\pi) \pm i\, m_y (\pi)$.
Here, ${\bi m} (\pi) = {\bi m}_{d-} - {\bi m}_{d+}$
is the hidden magnetic moment.
The propagator takes the form
\begin{equation}
D({\bi q},\omega) = {(2 s_1)^2\over{\chi_{\perp}}} 
[\omega^2 - \omega_b^2({\bi q})]^{-1}
\label{D}
\end{equation}
in the case of the above mean-field theory,
as well as in the case of 
the linear spin-wave approximation of the Heisenberg model (\ref{hsnbrg}) \cite{jpr_10}.
It shows a pole in frequency that disperses
acoustically as  $\omega_b({\bar{\bi q}}) = c_0 |{\bi q}|$
about ${\bi Q}_{\rm AF}$, where $c_0 = (\rho_s/\chi_{\perp})^{1/2}$
is the hidden-spin-wave velocity,
and where ${\bar{\bi q}} = {\bi q} + {\bi Q}_{\rm AF}$.
In the former case, $s_1$ is equal to the sub-lattice
magnetization,  $\langle m_{0,0}\rangle$,
while $s_1$ is given by the electron spin in the latter case.
Last, $\chi_{\perp}$ is the transverse spin susceptibility
of the hidden N\'eel state.

The previous mean-field theory implies that the hidden spinwaves in question
interact with independent electrons governed by the hopping Hamiltonian, $H_{\rm hop}$.
This is evident from the mean-field form of the interaction (\ref{U(pi)}) in isotropic form:
$-\sum_i \sum_{\alpha} U(\pi) {\bi m}_{i,\alpha} \cdot
2{\bi S}_{i,\alpha}$.
The transverse contributions yield the interaction 
$-\sum_i \sum_{\alpha} U(\pi) (m_{i,\alpha}^+ 
S_{i,\alpha}^- + m_{i,\alpha}^- S_{i,\alpha}^+$).
Plugging in expression (\ref{ci}) and its conjugate
for the electron creation and destruction  operators
yields the following interaction contribution to the Hamiltonian with hidden spinwaves:
\begin{eqnarray}
H_{\rm e-hsw} = -{1\over{\sqrt{2}}}{U(\pi)\over{a {\cal N}^{1/2}}}
 \sum_{\bi k} \sum_{{\bi k}^{\prime}}\sum_n
[m^+(\pi,{\bi q}) 
c_{\downarrow}^{\dagger}({\bar n},{\bar{\bi k}}^{\prime})
c_{\uparrow}(n,{\bi k}) & {\cal M}_{n,{\bi k};{\bar n},{\bar{\bi k}}^{\prime}}\nonumber\\
&+{\rm h.c.}],
\label{e-hsw}
\end{eqnarray}
where ${\bi q} = {\bi k}-{\bar{\bi k}}^{\prime}$ is the momentum transfer,
and where the matrix element above is the prior one for hidden order ($m=1$). 
(See \ref{ppndx_a}.)
Above, intra-band transitions are neglected because they do not show nesting.
Because we shall use Nambu-Gorkov formalism\cite{schrieffer_64,nambu_60,gorkov_58} below,
it is useful to write the above electron-hidden-spinwave interaction 
in terms of spinors:
\begin{eqnarray}
H_{\rm e-hsw} = \mp{1\over{\sqrt{2}}}{U(\pi)\over{{a \cal N}^{1/2}}}
 \sum_{\bi k} \sum_{{\bi k}^{\prime}}
[m^+(\pi,{\bi q})
{C}_{\downarrow}^{\dagger}({\bi k}^{\prime})
\tau_1
{C}_{\uparrow}({\bi k})
 & \sin[\delta({\bi k}) + \delta({\bi k}^{\prime})]\nonumber\\
&+{\rm h.c.}],
\label{E-hSW}
\end{eqnarray}
with spinor
\begin{equation}
C_s({\bi k}) =
\left[ {\begin{array}{c}
c_{s}(2,{\bi k}) \\ c_{s}(1,{\bar{\bi k}})
\end{array} } \right].
\label{spinor}
\end{equation}
Above, $\tau_1$ is the Pauli matrix along the $x$ axis.
Also, the explicit matrix
element ${\cal M}_{n,{\bi k};{\bar n},{\bar{\bi k}}^{\prime}}$ 
has been substituted in. (See \ref{ppndx_a}.)

\subsection{Electron propagator and Eliashberg equations}
Let $C_s({\bi k},t)$ denote the time evolution of 
the destruction operators (\ref{spinor}) $C_s({\bi k})$,
and let $C_s^{\dagger}({\bi k},t)$ denote the time evolution
for the conjugate creation operators $C_s^{\dagger}({\bi k})$.
The Nambu-Gorkov electron propagator is then
the Fourier transform 
$i G_s({\bi k},\omega) = \int d t_{1,2} e^{i \omega t_{1,2}}
\langle T[C_s({\bi k},t_1) C_s^{\dagger}({\bi k},t_2)]\rangle$,
where $t_{1,2} = t_1 - t_2$, and where $T$ is the time-ordering operator.
It is a $2 \times 2$  matrix. 
In the absence of interactions, its matrix inverse is then given by
\begin{equation}
G_0^{-1}({\bi k},\omega) =
\omega\, \tau_0 - \varepsilon_+({\bi k})\, \tau_3,
\label{1/G0}
\end{equation}
where $\tau_0$ is the $2 \times 2$ identity matrix,
and where $\tau_3$ is the Pauli matrix along the $z$ axis.
Guided by the previous mean field theory, let us next assume that
the matrix inverse of the Nambu-Gorkov Greens function takes the form
\begin{equation}
G_s^{-1}({\bi k},\omega) =
Z({\bi k},\omega)  \omega\, \tau_0 
- [\varepsilon_+({\bi k})-\nu]\, \tau_3
-Z({\bi k},\omega) ({\rm sgn}\, s) \Delta({\bi k})\,\tau_1.
\label{1/G}
\end{equation}
Here, $Z({\bi k},\omega)$ is the wavefunction renormalization,
$\Delta({\bi k})$ is the quasi-particle gap (\ref{gap}),
and $\nu$ is a relative energy shift of the bands that preserves perfect nesting.
In particular,
the form (\ref{1/G}) of the Nambu-Gorkov Greens function
is consistent with the  perfect nesting condition
$\varepsilon_{\pm}({\bi k}+{\bi Q}_{\rm AF})\mp\nu = -
[\varepsilon_{\mp}({\bi k})\pm\nu]$
that is equivalent to  (\ref{prfct_nstng}).
Matrix inversion of (\ref{1/G}) yields the Nambu-Gorkov Greens function\cite{schrieffer_64,nambu_60,gorkov_58}
$G = \sum_{\mu = 0}^{3} G^{(\mu)} \tau_{\mu}$, with components
\begin{eqnarray}
G_s^{(0)} &=& {1\over{2 Z}}
\Biggl({1\over{\omega-E}} + {1\over{\omega+E}}\Biggr),\nonumber \\
G_s^{(1)} &=& {1\over{2 Z}}
\Biggl({1\over{\omega-E}} - {1\over{\omega+E}}\Biggr)
{\Delta\over E} ({\rm sgn}\, s), \nonumber \\
G_s^{(3)} &=& {1\over{2 Z}}
\Biggl({1\over{\omega-E}} - {1\over{\omega+E}}\Biggr)
{(\varepsilon_+-\nu)\over Z E},
\label{G}
\end{eqnarray}
and $G_s^{(2)} = 0$. Above, the excitation energy is
\begin{equation}
E({\bi k},\omega) = 
\sqrt{\Biggl[{\varepsilon_+({\bi k})-\nu\over{Z({\bi k},\omega)}}\Biggr]^2 
+ \Delta^2({\bi k})}.
\label{E_k}
\end{equation}

To obtain the Eliashberg equations,
recall first the definition of the self-energy correction:
$G^{-1} = G_0^{-1} - \Sigma$.
Comparison of the inverse Greens functions
(\ref{1/G0}) and (\ref{1/G})
then yields\cite{schrieffer_64,scalapino_69}
\begin{equation}
\Sigma_s({\bi k},\omega) =
[1-Z({\bi k},\omega)] \omega\, \tau_0 
-\nu\, \tau_3
+Z({\bi k},\omega) ({\rm sgn}\, s) \Delta({\bi k})\,\tau_1
\label{Sigma}
\end{equation}
for it.
Next, we neglect vertex corrections from the electron-hidden-spinwave interaction (\ref{E-hSW}).
This approximation will be justified {\it a posteriori} at the end of the next subsection.
The self-energy correction is then approximated by
\begin{eqnarray}
\Sigma_s({\bi k},i\omega_n) = - k_B T
\int {d^2 k^{\prime}\over{(2\pi)^2}}  \sum_{i\omega_{n^{\prime}}}
& {U^2(\pi)\over 2} \sin^2[\delta({\bi k})+\delta({\bi k}^{\prime})] \cdot \nonumber\\
& \cdot D({\bi q},i\omega_m) \tau_1 G_{\bar s}({\bi k}^{\prime},i\omega_{n^{\prime}})\tau_1,
\label{self-energy}
\end{eqnarray}
with $i\omega_{m} = i \omega_n - i \omega_{n^{\prime}}$,
and with ${\bi q} = {\bi k} - {\bar{\bi k}}^{\prime}$.
Here, we have Wick rotated to pure imaginary Matsubara frequencies
at non-zero temperature $T$. Observe, finally, that
$\tau_1 \tau_{\mu} \tau_1 = {\rm sgn}_{\mu}\tau_{\mu}$,
where ${\rm sgn}_0 = +1 = {\rm sgn}_1$, and 
where ${\rm sgn}_2 = -1 = {\rm sgn}_3$.
Identifying expressions (\ref{Sigma}) and (\ref{self-energy}) for 
the self-energy corrections then yields the following self-consistent
Eliashberg equations at non-zero temperature:
\begin{eqnarray}
-[Z({\bi k},i\omega_n)-1] i\omega_n &= - k_B T
\int {d^2 k^{\prime}\over{(2\pi)^2}} \sum_{i\omega_{n^{\prime}}}
& {U^2(\pi)\over 2}  \sin^2[\delta({\bi k})+\delta({\bi k}^{\prime})] \cdot \nonumber\\
& & \cdot D({\bi q},i\omega_m) G_{\bar s}^{(0)}({\bi k}^{\prime},i\omega_{n^{\prime}}), \nonumber\\
\qquad\qquad -\nu &= + k_B T
\int {d^2 k^{\prime}\over{(2\pi)^2}} \sum_{i\omega_{n^{\prime}}}
& {U^2(\pi)\over 2}  \sin^2[\delta({\bi k})+\delta({\bi k}^{\prime})] \cdot \nonumber\\
& & \cdot D({\bi q},i\omega_m) G_{\bar s}^{(3)}({\bi k}^{\prime},i\omega_{n^{\prime}}), \nonumber\\
\qquad Z({\bi k},i\omega_n) ({\rm sgn}\, s) \Delta({\bi k}) &= - k_B T
\int {d^2 k^{\prime}\over{(2\pi)^2}} \sum_{i\omega_{n^{\prime}}}
& {U^2(\pi)\over 2}  \sin^2[\delta({\bi k})+\delta({\bi k}^{\prime})] \cdot \nonumber\\
& & \cdot D({\bi q},i\omega_m) G_{\bar s}^{(1)}({\bi k}^{\prime},i\omega_{n^{\prime}}).\nonumber \\
\label{E_eqs_T}
\end{eqnarray}
The Greens functions above are listed in (\ref{G}).

Last,
the above Eliashberg equations can be expressed  at real frequency.  
In particular, it becomes useful to
write the propagator for hidden spinwaves (\ref{D}) as
\begin{equation}
D({\bi q},i\omega_m) = {(2 s_1)^2\over{\chi_{\perp}}} 
{1\over{2\omega_b({\bi q})}}\Biggl[{1\over{i\omega_m - \omega_b({\bi q})}}
-{1\over{i\omega_m + \omega_b({\bi q})}}\Biggr].
\label{d}
\end{equation}
A series of  decompositions into partial fractions followed by summations 
of Matsubara frequencies
yields Eliashberg equations
in terms of Fermi-Dirac
and Bose-Einstein distribution functions at real frequency. 
They are listed in \ref{ppndx_c}.
At zero temperature, these reduce to
\begin{eqnarray}
[Z({\bi k},\omega)-1] \omega &=& 
\int {d^2 k^{\prime}\over{(2\pi)^2}} 
 U^2(\pi) {s_1^2\over{\chi_{\perp}}} {\sin^2[\delta({\bi k})+\delta({\bi k}^{\prime})]\over{Z({\bi k}^{\prime},\omega)}} \cdot \nonumber\\
&& \cdot {1\over{2\omega_b({\bi q})}}
\Biggl[{1\over{\omega_b({\bi q})+E({\bi k}^{\prime})-\omega}}
-{1\over{\omega_b({\bi q})+E({\bi k}^{\prime})+\omega}}\Biggr], \nonumber\\
-\nu &=& 
\int {d^2 k^{\prime}\over{(2\pi)^2}}
U^2(\pi) {s_1^2\over{\chi_{\perp}}} {\sin^2[\delta({\bi k})+\delta({\bi k}^{\prime})]\over{Z({\bi k}^{\prime},\omega)}}
{\varepsilon_+({\bi k}^{\prime})-\nu\over{Z({\bi k}^{\prime},\omega) E({\bi k}^{\prime})}} \cdot \nonumber\\
&& \cdot {1\over{2\omega_b({\bi q})}}
\Biggl[{1\over{\omega_b({\bi q})+E({\bi k}^{\prime})-\omega}}
+{1\over{\omega_b({\bi q})+E({\bi k}^{\prime})+\omega}}\Biggr], \nonumber\\
 Z({\bi k},\omega) \Delta({\bi k}) &=& 
 \int {d^2 k^{\prime}\over{(2\pi)^2}}
U^2(\pi) {s_1^2\over{\chi_{\perp}}} {\sin^2[\delta({\bi k})+\delta({\bi k}^{\prime})]\over{Z({\bi k}^{\prime},\omega)}} 
{\Delta({\bi k}^{\prime})\over{E({\bi k}^{\prime})}} \cdot \nonumber\\
&& \cdot {1\over{2\omega_b({\bi q})}}
\Biggl[{1\over{\omega_b({\bi q})+E({\bi k}^{\prime})-\omega}}
+{1\over{\omega_b({\bi q})+E({\bi k}^{\prime})+\omega}}\Biggr]. \nonumber\\
\label{E_eqs}
\end{eqnarray}
Below, we find solutions to the above equations.

\subsection{Fermi-surface pockets at corner of Brillouin zone}
The central aim of this paper is to reveal a Lifshitz transition
from the Fermi surface depicted by Fig. \ref{FS0}
to electron/hole pockets at the corner of the two-iron Brillouin zone.
Let us therefore work in the normal state
and take the trivial solution for the gap equation (\ref{E_eqs}): $\Delta({\bi k}) = 0$.
Furthermore, let us neglect any angular dependence acquired either
by the wavefunction renormalization, $Z({\bi k},\omega)$, or by the relative 
energy shift of the bands, $\nu$, on momentum around the Fermi surface:
$\varepsilon_+({\bi k}) = \nu$.
This is exact for $\nu$
near the upper band edge of $\varepsilon_+({\bi k})$
in the absence of nearest-neighbor intra-orbital hopping, 
$t_1^{\parallel} = 0$,
in which case circular Fermi surface pockets exist at $(\pi/a,0)$ and at $(0,\pi/a)$.
Following the standard procedure\cite{scalapino_69},
we then multiply both sides of the remaining two Eliashberg equations (\ref{E_eqs})
by $\delta[\varepsilon_+({\bi k}) - \nu] / D_+(\nu)$
and integrate in momentum over the first Brillouin zone.
The previous Eliashberg equations (\ref{E_eqs}) thereby reduce to
\numparts
\begin{eqnarray}
(Z-1) \omega &=& 
\int_{-W_{\rm bottom}}^{+W_{\rm top}} d\Biggl({\varepsilon^{\prime}\over{Z}}\Biggr)
\int_0^{\infty} d\Omega\,  U^2 F_0(\Omega;\nu,\nu)\cdot \nonumber\\
&& \cdot {1\over{2}}
\Biggl[{1\over{\Omega+|\varepsilon^{\prime}-\nu|/Z-\omega}}
-{1\over{\Omega+|\varepsilon^{\prime}-\nu|/Z+\omega}}\Biggr],
\label{2_E_eqs_a} \\
\qquad -\nu &=& 
\int_{-W_{\rm bottom}}^{+W_{\rm top}} d\Biggl({\varepsilon^{\prime}\over{Z}}\Biggr)
\int_0^{\infty} d\Omega\,  U^2 F_0(\Omega;\nu,\nu)
{\varepsilon^{\prime}-\nu
\over{|\varepsilon^{\prime}-\nu|}}\cdot \nonumber\\
&& \cdot {1\over{2}}
\Biggl[{1\over{\Omega+|\varepsilon^{\prime}-\nu|/Z-\omega}}
+{1\over{\Omega+|\varepsilon^{\prime}-\nu|/Z+\omega}}\Biggr],
\label{2_E_eqs_b}
\end{eqnarray}
\endnumparts
where
\begin{eqnarray}
U^2 F_0(\Omega;\varepsilon,\varepsilon^{\prime}) = {1\over{D_+(\varepsilon)}}
\int {d^2 k\over{(2\pi)^2}} & \int {d^2 k^{\prime}\over{(2\pi)^2}}
 U^2(\pi) {s_1^2\over{\chi_{\perp}}}
{\sin^2[\delta({\bi k})+\delta({\bi k}^{\prime})]
\over{\omega_b({\bi q})}} \cdot \nonumber \\
& \cdot \delta[\varepsilon_+({\bi k})-\varepsilon] \delta[\varepsilon_+({\bi k}^{\prime})-\varepsilon^{\prime}]
\delta[\omega_b({\bi q}) - \Omega],\nonumber \\
\label{U2F}
\end{eqnarray}
and where the wavefunction renormalization is averaged over the new Fermi surface:
$Z({\bi k,\omega})\rightarrow
[D_+(\nu)]^{-1} (2\pi)^{-2} \int_{\rm BZ} d^2 k\, Z({\bi k},\omega) \delta[\varepsilon_+({\bi k})-\nu]$.
Above, we have also approximated 
the function $U^2 F_0(\Omega;\nu,\varepsilon^{\prime})$ of $\varepsilon^{\prime}$
by its value at the renormalized chemical potential, $U^2 F_0(\Omega;\nu,\nu)$.
It is also understood in (\ref{U2F})
that the limit implicit in the last $\delta$-function factor
is taken last.

\begin{figure}
\includegraphics{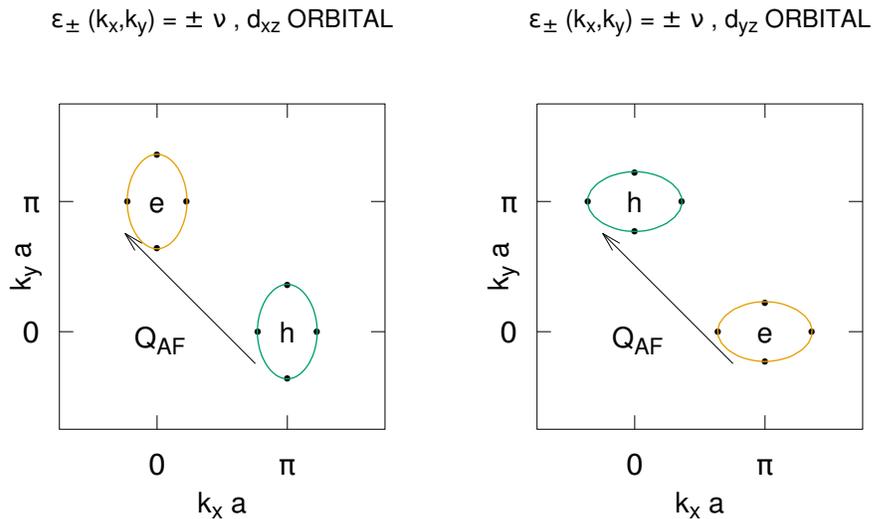}
\caption{Shown are the the renormalized Fermi surface pockets for inter-band energy shift
$\nu$ near the upper edge of the bonding band $\varepsilon_+({\bi k})$,
and for $-\nu$
near the lower edge of the anti-bonding band $\varepsilon_-({\bi k})$.
The orbital character is only approximate, 
although it becomes exact as the area of the Fermi surface pockets vanishes as $U_0$ diverges.}
\label{FS1}
\end{figure}

The effective spectral weight of the hidden spinwaves, $U^2 F_0(\Omega;\nu,\nu)$,
can be evaluated by choosing coordinates for the momentum of the electron,
 $(k_{\parallel}, k_{\perp})$,
that are respectively  parallel and perpendicular
to the Fermi surface of the bonding band (FS$_+$):
 $\nu = \varepsilon_+({\bi k})$.
(See Figs. \ref{FS0} and \ref{FS1}.)
This yields the intermediate result
\begin{eqnarray}
U^2 F_0(\Omega;\nu,\nu) = {1\over{D_+(\nu)}}
\oint_{{\rm FS}_+} {d k_{\parallel}\over{(2\pi)^2}} 
& \oint_{{\rm FS}_+} {d k_{\parallel}^{\prime}\over{(2\pi)^2}}
 U^2(\pi) {s_1^2\over{\chi_{\perp}}}
{1 \over{\Omega}} \cdot \nonumber \\
& \cdot {\sin^2[\delta({\bi k})+\delta({\bi k}^{\prime})]\over{|{\bi v}({\bi k})| |{\bi v}({\bi k}^{\prime})|}}
\delta[\omega_b({\bi q}) - \Omega],
\end{eqnarray}
where ${\bi v} = \partial\varepsilon_+ /\partial {\bi k}$ is the group velocity.
Yet 
the dispersion of the spectrum of hidden spinwaves
$\omega_b({\bar{\bi q}})$
is acoustic at low energy.
This then yields the following dependence on frequency
for their effective spectral weight:
$U^2 F_0(\Omega;\nu,\nu) = \epsilon_{\rm E}(\nu)/\Omega$ as $\Omega\rightarrow 0$,
with a constant pre-factor
\begin{equation}
\epsilon_{\rm E}(\nu) = {1\over{D_+(\nu)}}
\oint_{{\rm FS}_+} {d k_{\parallel}\over{(2\pi)^4}} 
 U^2(\pi) {s_1^2\over{\chi_{\perp}}}
{[\sin\, 2\delta({\bi k})]^2\over{c_0 |{\bi v}({\bi k})|^2}}.
\label{epsilon_E}
\end{equation}
Above, $c_0$ is the velocity of hidden spinwaves at ${\bi Q}_{\rm AF}$.

We can now find solutions to the remaining Eliashberg equations
 (\ref{2_E_eqs_a}) and (\ref{2_E_eqs_b}).
In particular,
assume that the relative energy shift $\nu$ lies near the upper edge
of the bonding band $\varepsilon_+({\bi k})$
at $(\pi/a,0)$ and at $(0,\pi/a)$.
Figure \ref{FS1} displays the Fermi surfaces in such case.
Substituting in the simple pole in frequency above for $U^2 F_0(\Omega;\nu,\nu)$ yields
the first Eliashberg equation:
\begin{eqnarray}
\omega(Z-1) = {\epsilon_{\rm E}\over 2}
\int_0^{\omega_{\rm uv}} {d\Omega\over{\Omega}}
{\rm ln}\Biggl|{\Omega + \omega\over{\Omega - \omega}} \cdot
{W/Z + \Omega-\omega\over{W/Z + \Omega+\omega}}\Biggr|.
\label{1st_E_eq}
\end{eqnarray}
Here,
we have reversed the order of integration.
Also above, $[\nu - W,\nu]$ is the range of integration over
$\varepsilon^{\prime}$  in (\ref{2_E_eqs_a}),
where $W=W_{\rm bottom} + W_{\rm top}$ is the bandwidth of $\varepsilon_+ ({\bi k})$,
while $\omega_{\rm uv}$ is an ultra-violet cutoff in frequency for the hidden spinwaves.
Assuming $W / Z \gg \omega$ yields the equation 
\begin{eqnarray}
\omega(Z-1) &=& {\epsilon_{\rm E}\over 2}
\int_0^{\infty} d x {1\over x}
{\rm ln}\Biggl|{x+1\over{x-1}}\Biggr| \nonumber \\
&=& {3\over 2} \zeta(2)  \epsilon_{\rm E}
\label{1st_E_e}
\end{eqnarray}
in the low-frequency limit, where $x = \Omega/\omega$ above.
The final result for the wavefunction renormalization at the Fermi surface is then
$Z = (\pi^2 / 4) (\epsilon_{\rm E} / \omega)$
as $\omega\rightarrow 0$.
By (\ref{G}),
the spectral weight of quasi-particle excitations is $1/Z$.
It therefore vanishes at the Fermi level, $\omega = 0$.
This result is then consistent with the characterization of the hSDW state as a Mott insulator.

The second Eliashberg equation (\ref{2_E_eqs_b}) can be evaluated in a similar way.
After substituting in the simple pole in frequency for $U^2 F_0(\Omega;\nu,\nu)$,
integrating first over $\varepsilon^{\prime}$ 
yields the equation
\begin{eqnarray}
\nu = {\epsilon_{\rm E}\over 2} \int_0^{\omega_{\rm uv}} {d\Omega\over{\Omega}}
{\rm ln}\Biggl|
{(W/Z + \Omega)^2 - \omega^2\over{\Omega^2-\omega^2}}
\Biggr|. \nonumber
\end{eqnarray}
Assuming, once again, the inequality
$W / Z \gg \omega$
then yields the following equation
in the low-frequency limit:
\begin{eqnarray}
\nu = \epsilon_{\rm E} \int_0^{\omega_{\rm uv}} {d\Omega\over{\Omega}}
{\rm ln}\Biggl|
{W\over{\omega Z}}
\Biggr|. \nonumber
\end{eqnarray}
Here, we have expanded the previous argument
of the logarithm in powers of $\Omega$.
The final result for the second Eliashberg equation is then
\begin{eqnarray}
{\rm lim}_{\omega\rightarrow 0}\nu = 
\epsilon_{\rm E}\, {\rm ln} 
\Biggl( {\omega_{\rm uv}\over{\omega_{\rm ir}}} \Biggr)
{\rm ln}
\Biggl( {4\, W\over{\pi^2 \epsilon_{\rm E}}} \Biggr) , 
\label{nu}
\end{eqnarray}
where $\omega_{\rm ir} \sim c_0/L$ is an infra-red cutoff in frequency.
Above, the previous result from the first Eliashberg equation has been substituted in.
We can now check the previous inequality that was assumed.
The second Eliashberg equation (\ref{nu}) implies that 
the energy scale (\ref{epsilon_E})
is of order
$\epsilon_{\rm E} \sim \varepsilon_+(\pi/a,0) / {\rm ln}(\omega_{\rm uv}/\omega_{\rm ir})$
for $\nu$ near the upper edge of the bonding band $\varepsilon_+({\bi k})$.
The ratio
$W/\omega Z  = 4 W /\pi^2 \epsilon_{\rm E}$
is therefore of order ${\rm ln}(\omega_{\rm uv}/\omega_{\rm ir})$,
which {\it diverges} logarithmically as the infra-red cutoff in
frequency $\omega_{\rm ir}$ tends to zero.

We shall finally estimate the Eliashberg energy scale (\ref{epsilon_E})
$\epsilon_{\rm E}$.  For simplicity, assume small circular Fermi surface
pockets ($t_1^{\parallel} = 0$) of Fermi radius $k_F$,
which is related to the concentration $x_0$ of electron/holes per pocket 
by $k_F a = (2\pi x_0)^{1/2}$.
The Fermi velocity is
then $v_F = 2 t_1^{\perp} k_F a^2$.
Also, the phase shift (\ref{s_2dlt}) is approximately
$\sin\, 2\delta ({\bi k}) =  
[(t_2^{\perp}/i)/2 t_1^{\perp}] (k_F a)^2\, \sin\, 2\phi$,
where $\phi$ is the angle that ${\bi k}$ makes about the center of
the Fermi surface pocket at $(\pi/a,0)$ or at $(0,\pi/a)$.
We then get the expression
\begin{equation}
\epsilon_{\rm E} = {1\over 16} \Biggl({x_0\over{2\pi}}\Biggr)^{3/2}
{U^2(\pi)\over{a^2 D_+(\nu)}} {s_1^2\over{a^2 \chi_{\perp}}}
{|t_2^{\perp}|^2\over{(c_0/a) |t_1^{\perp}|^4}}
\label{est_eps_e}
\end{equation}
for the Eliashberg energy scale (\ref{epsilon_E}).
Comparing this estimate with the second Eliashberg equation (\ref{nu}),
while fixing $\nu$ to 
the upper edge of the band ${\varepsilon}_+({\bi k})$,
then yields that the area of the electron/hole Fermi surface pockets
shown in Fig. \ref{FS1}
vanishes logarithmically with the size $L$ of the system
for {\it any} positive  $U(\pi)$.
Note, however,
that such behavior is effectively ruled out at the weak-coupling limit,
$U(\pi)\rightarrow 0$,
because the ordered magnetic moment $s_1$ vanishes
exponentially in such case.
On the other hand,
if instead the scale $L$ of the system is fixed,
then the previous comparison of (\ref{nu}) and (\ref{est_eps_e})
yields $U(\pi)\propto x_0^{-3/4}$. By the previous estimate for the phase shift
at the new Fermi surface pockets, 
the electron-hidden-spin-wave interaction (\ref{E-hSW})
then scales as $U(\pi)^{-1/3}$.
This justifies our neglect of vertex corrections
at the limit of strong on-site
repulsion, $U_0\rightarrow\infty$.

Last, a self-consistent solution to the above Eliashberg equations at the
limit of large on-iron-site-orbital Coulomb repulsion, $U_0$,
also exists at a relative shift of the bands $\nu$ near the bottom edge of the
bonding band, $\varepsilon_+({\bi k})$, instead.
In particular,
$[\nu, \nu + W]$ is now the range of integration over
$\varepsilon^{\prime}$  in
the Eliashberg equations (\ref{2_E_eqs_a}) and (\ref{2_E_eqs_b}).
The previous results for the wavefunction renormalization (\ref{1st_E_e})
and for the relative energy shift between the two bands (\ref{nu}) 
hold after making the replacement $\nu = W_{\rm top}$ with $-\nu = W_{\rm bottom}$ 
in the latter.
It is important, now, to observe that the density of states of the bonding band
at the upper band edge is larger than the density of states at the bottom edge
by Fig. \ref{DoS}.
The condensation energy is of order $-D_+(\nu) \Delta_0^2$,
however. By the definition (\ref{Delta0}) for $\Delta_0$
and by Fig. \ref{magnetic_moment} for the ordered magnetic moment,
the condensation energy
dominates the kinetic (hopping) energy at strong on-site repulsion $U_0$
compared to the bandwidth. This argues in favor of the former solution
in such case, with $\nu$ at 
the upper edge of the band $\varepsilon_+({\bi k})$.

\section{Discussion} 
The previous mean field theory analysis of the extended two-orbital
Hubbard model for heavily electron-doped FeSe finds that hidden N\'eel antiferromagnetic order
is expected at perfect nesting (\ref{prfct_nstng}) $t_2^{\parallel} =0$
when true N\'eel order is suppressed by magnetic frustration\cite{jpr_ehr_09,jpr_10}.
(See Figs. \ref{FS0} and \ref{FS1}.)
Below, we compare the observable consequences that have been listed above 
with analogous theoretical results at the strong-coupling limit\cite{jpr_17}, 
$U_0 \rightarrow \infty$,
and with recent experimental evidence for such hidden magnetic order
in the superconducting state of intercalated FeSe\cite{davies_16,pan_17,ma_17}.
We also argue why the effects of the iron $3 d_{xy}$ orbital can be neglected.

\subsection{Comparison of weak coupling and strong coupling}
In subsections \ref{hydro} and \ref{rho_chi}, 
we showed how mean field theory for the hidden N\'eel state
of the two-orbital Hubbard model that describes
heavily electron-doped FeSe
agrees both qualitatively and quantitatively with the corresponding Heisenberg model
at large on-iron-site-orbital repulsion.
In particular, a hydrodynamical analysis (\ref{dynamics})
predicts two acoustically dispersing 
spin-wave excitations
per momentum near the ``checkerboard'' wavenumber ${\bi Q}_{\rm AF}$.
This agrees with the large-$s_0$ analysis of 
the corresponding Heisenberg model\cite{jpr_10}.
Second,
the transverse spin susceptibility of the hSDW state (\ref{RPA})
was calculated above as well.  
At weak hybridization between the $3 d_{xz}$ and $3 d_{yz}$ orbitals,
the transverse susceptibility
of the hSDW state at the limit of strong on-iron-site-orbital Coulomb repulsion
(\ref{strng_cplng}) is found to agree with the same quantity calculated
from the corresponding Heisenberg model
in the large-$s_0$ limit\cite{jpr_10}.
Also,
the spin rigidity (\ref{rho_s}) 
of the hSDW state
was computed above at the limit of strong on-iron-site-orbital repulsion.
A comparison with the same results for the corresponding Heisenberg model (\ref{hsnbrg})
yields exchange coupling constants that are consistent with hidden N\'eel order.
(See the Goldstone mode in Fig. \ref{spn_spctrm}b.)

Figure \ref{FS1} is the central result of the paper, however.
It shows the Fermi surfaces of the extended Hubbard model
in the hSDW state at half-filling
and at strong on-iron-site-orbital Coulomb repulsion,
as predicted by Eliashberg theory in the particle-hole channel.
The rigid-band approximation, in turn, predicts electron-type Fermi surface pockets alone
at wavenumbers $(\pi/a,0)$ and $(0,\pi/a)$ upon electron doping at concentrations per pocket $x > x_0$.
Here, $x_0$ denotes the concentration of electrons/holes inside the Fermi surface pockets
shown in Fig. \ref{FS1}.
It vanishes as $U_0$ diverges.
This argument agrees with Schwinger-boson-slave-fermion mean field theory
of the corresponding local-moment ($t$-$J$) model at electron doping\cite{jpr_17},
in which case $U_0\rightarrow\infty$ and $x_0\rightarrow 0$,
and in which case only the electron-type Fermi surface pockets 
shown in Fig. \ref{FS1} appear.
It also notably agrees with ARPES on heavily electron-doped FeSe\cite{liu_12,zhou_13,zhao_16,niu_15}.
In particular, $x_0$ may represent a threshold concentration of electron doping
at which hSDW order gives way to superconductivity.

\subsection{Comparison of hidden magnetic order with experiment}
\begin{figure}
\includegraphics{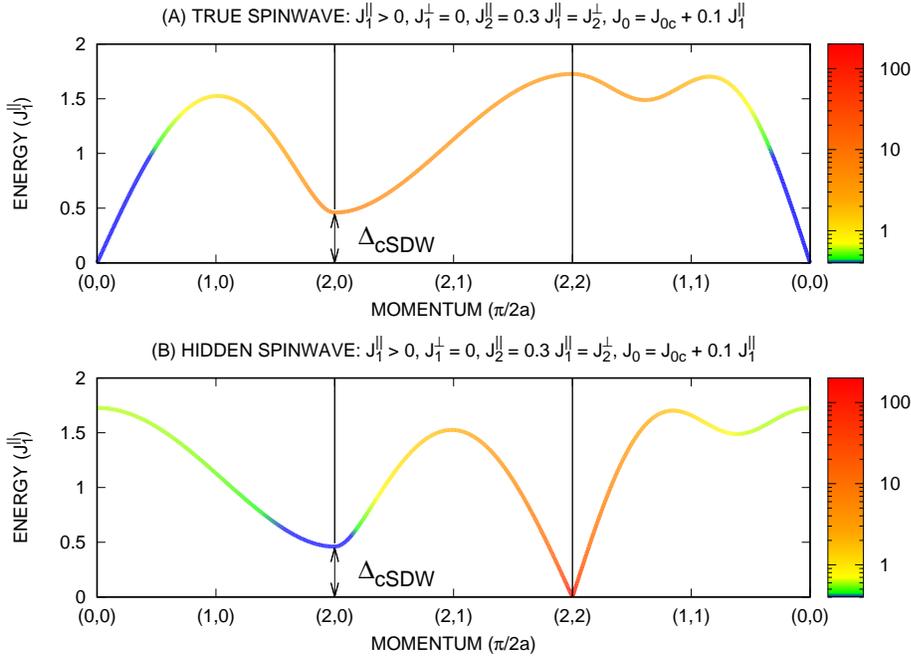}
\caption{Spin-excitation spectrum for hidden spin-density-wave state 
of the Heisenberg model (\ref{hsnbrg}) at the large-$s_0$ limit,
but with the replacement
$J_1^{\perp}\rightarrow J_1^{\perp} -  t_1^{\perp} x /(1-x)^2 s_0$.
(See refs. \cite{jpr_17,jpr_10} and the main text.)
Here, $s_0 = 1/2$, $t_1^{\perp} = 5\, J_1^{\parallel}$ and $x = 0.01$.}
\label{spn_spctrm}
\end{figure}

The local-moment limit of the present extended Hubbard model for heavily electron-doped FeSe 
is achieved at strong on-site-orbital repulsion, $U_0\rightarrow\infty$.
At half filling,
it results in a two-orbital Heisenberg model over the square lattice of the form\cite{jpr_10}
\begin{eqnarray}
H_{\rm Hsnbrg} = 
& \sum_{\langle i,j\rangle}\sum_{\alpha} (J_1^{\parallel} {\bi S}_{i,\alpha}\cdot{\bi S_{j,\alpha}} +
J_1^{\perp} {\bi S}_{i,\alpha}\cdot{\bi S}_{j,{\bar\alpha}}) \nonumber \\
& + \sum_{\langle\langle i,j\rangle\rangle}\sum_{\alpha} (J_2^{\parallel} {\bi S}_{i,\alpha}\cdot{\bi S_{j,\alpha}} +
J_2^{\perp} {\bi S}_{i,\alpha}\cdot{\bi S}_{j,\bar{\alpha}}),
\label{hsnbrg}
\end{eqnarray}
where $\alpha = d-$ or $d+$.
In particular,
the results obtained in subsection \ref{rho_chi} 
for the transverse spin susceptibility
and for the spin rigidity
of the hSDW state
are consistent with the following assignments for the Heisenberg exchange coupling constants:
$J_1^{\parallel} = J_1^{({\rm sprx})}$,
$J_1^{\perp} = J_1^{({\rm sprx})} -  \rho_s / 2 s_0^2$,
and $J_2^{\parallel} = J_2^{({\rm sprx})} = J_2^{\perp}$.
Here, $\rho_s > 0$ is the spin rigidity  of the hSDW (\ref{RHO_S}).
Adding electrons at this strong-coupling limit
can be studied analytically within the Schwinger-boson-slave-fermion formulation
when only inter-orbital nearest neighbor hopping, $t_1^{\perp} > 0$, exists\cite{jpr_17}.
The mean field theory of the corresponding hSDW state is well behaved in such case.
As mentioned previously, it shows two electron-type Fermi surface pockets at the
corner of the two-iron Brillouin zone, with $3d_{xz}$ and $3d_{yz}$ orbital
character respectively. (Cf. Fig. \ref{FS1}.)
Schwinger-boson-slave-fermion mean field theory also finds two branches of spin-wave excitations
that correspond to true and to hidden magnetic moments,
${\bi S}_{i,d-} + {\bi S}_{i,d+}$ and ${\bi S}_{i,d-} - {\bi S}_{i,d+}$, respectively.
They are governed by the Heisenberg model (\ref{hsnbrg}) in the large-$s_0$ limit,
but with the replacement
$J_1^{\perp}\rightarrow J_1^{\perp} -  t_1^{\perp} x /(1-x)^2 s_0$.
Here, $s_0$ is the electron spin, and $x$ is the electron doping concentration per site-orbital.
Figure \ref{spn_spctrm} shows the spin-wave spectra from such a large-$s_0$ approximation
for the hSDW state of the local-moment model
near a critical Hund's Rule coupling $J_{0c}$ where the spectrum softens completely
at ``stripe'' SDW wavenumbers $(\pi/a,0)$ and $(0,\pi/a)$ \cite{jpr_17}; 
i.e., $\Delta_{\rm cSDW}\rightarrow 0$.

\begin{figure}
\includegraphics{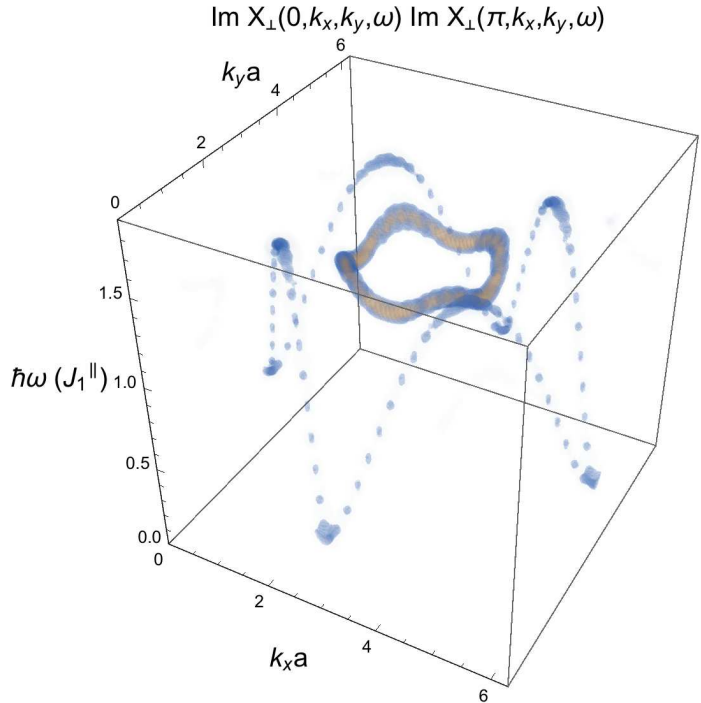}
\caption{Curves in momentum versus energy where the dispersion of true spinwaves is degenerate with 
that of hidden spinwaves: $\omega_b(0,{\bi k}) = \omega_b(\pi,{\bi k})$. Parameters
for the Heisenberg model (\ref{hsnbrg}) are listed in Fig. \ref{spn_spctrm}.}
\label{fltng_rng}
\end{figure}

The results shown by Fig. \ref{spn_spctrm}
for the spin-excitation spectrum of the hSDW state 
are obtained from the local-moment model for
heavily electron-doped FeSe that includes only inter-orbital
nearest neighbor hoping, $t_1^{\perp} > 0$.
The $3d_{xz}$ and $3d_{yz}$ orbitals are good quantum numbers in such case.
In particular, they are respectively even and odd under
 swap of the $d-$ and the $d+$ orbitals, $P_{d,{\bar d}}$.
Likewise,
the true and hidden magnetic moments just cited
are  respectively even and odd under\cite{jpr_mana_pds_14} $P_{d,{\bar d}}$.
Unfortunately,
unlike the previous analysis of the extended Hubbard model,
the Schwinger-boson-slave-fermion mean field theory that such results are based on
is not well behaved when mixing between the two orbitals
(pure imaginary $t_2^{\perp}$)
is turned on\cite{jpr_17}.
Orbital swap $P_{d,{\bar d}}$ is no longer a global symmetry in such case.
Figure \ref{fltng_rng} shows, however,
the points in momentum and energy at which the two branches of the spin-excitation spectrum
are degenerate.  It reveals a ``floating ring'' of low-energy magnetic excitations
about the N\'eel wave number ${\bi Q}_{\rm AF}$.
Observable spin excitations should be brightest along the floating ring at 
weak mixing between the $3d_{xz}$ and $3d_{yz}$ orbitals.
Similar low-energy spin resonances around ${\bi Q}_{\rm AF}$
have been observed recently in the superconducting phase of
intercalated FeSe by inelastic neutron scattering\cite{davies_16,pan_17,ma_17}.
Such experiments are then consistent with the hSDW state proposed here
for heavily electron-doped FeSe.

\subsection{Iron $3 d_{xy}$ Orbital}
In addition to the iron $3 d_{xz}$ and $3 d_{yz}$
orbitals considered in the present extended Hubbard model,
ARPES on iron chalcogenides coupled with density-functional theory calculations 
indicate that the iron $3d_{xy}$ orbital also plays
an important role in the electronic structure\cite{yi_15}.
Without loss of generality, let us then simply add this orbital to
the four on-site terms in $H_U$ (\ref{U})
and to both super-exchange terms in $H_{\rm sprx}$ (\ref{sprx}).
Next, let us work within the approximation that no hybridization exists 
in between the $3 d_{xy}$ band, $\varepsilon_{xy}({\bi k})$,
and the $3 d_{xz} / 3 d_{yz}$ bands, $\varepsilon_-({\bi k})$ and $\varepsilon_+({\bi k})$.
Assume also that the $3 d_{xy}$ band shows no nesting.
Now recall that the net magnetic moment due to the $3 d_{xz} / 3 d_{yz}$ orbitals
is null in the hidden magnetic order state.
This implies that the paramagnetic state for electrons in the $3 d_{xy}$ orbital,
$\langle n_{i,d_{xy},\uparrow} \rangle = \langle n_{i,d_{xy},\downarrow} \rangle$,
is stable within the mean-field approximation outlined in subsection \ref{mft}.
The paramagnetic state of $3 d_{xy}$ electrons is thereby {\it decoupled}
from the $3 d_{xz} / 3 d_{yz}$ electrons in the hidden magnetic order state.
The former then acts as a potential charge reservoir for the latter.
What happens in the case where hybridization exists between all three orbitals\cite{Lee_Wen_08}
remains an open question that lies outside the scope of the present study.

\section{Summary and Conclusions}
Understanding the mechanism behind the high-temperature superconductivity
displayed by heavily electron-doped iron selenide remains elusive.
In an attempt to solve this mystery,
we have shown how the electron-type Fermi surface pockets that exist at
the corner of the two-iron Brillouin zone
in heavily electron-doped iron-selenide can emerge 
from an extended Hubbard model over the square lattice
of iron atoms that includes only the $3d_{xz}$ and $3d_{yz}$ orbitals.
At half-filling,
and in the absence of next-nearest neighbor intra-orbital hopping,
perfect nesting exists 
between hole-type and electron-type Fermi surfaces
displayed by Fig. \ref{FS0}.
The nesting wavenumber is $(\pi/a,\pi/a)$, 
which corresponds to checkerboard (N\'eel) order.
It notably differs from  parent compounds to iron-pnictide high-temperature superconductors,
which display ``stripe'' spin-density order, with nesting vector $(\pi/a,0)$.
The former checkerboard nesting
can lead to hidden N\'eel order 
that violates Hund's Rule
when true N\'eel order is suppressed
by magnetic frustration\cite{jpr_ehr_09,jpr_10}.
An extended Hartree-Fock calculation of the Eliashberg type reveals that hole and electron
Fermi surfaces become centered at the corner of the two-iron Brillouin zone
at moderate to strong on-site Coulomb repulsion
because of the exchange of antiferromagnetic spin fluctuations.
The electron/hole concentration $x_0$ that corresponds to the area of these
Fermi surface pockets vanishes as the on-site Coulomb repulsion diverges.
Sufficiently strong electron doping $x > x_0$ will then produce
a rigid shift of such a renormalized band structure,
with electron Fermi surface pockets alone
that are similar to those seen by ARPES in heavily electron-doped FeSe.

We have also shown that the extended two-orbital Hubbard model
leads to a local-moment model in the limit of strong on-site Coulomb repulsion
that harbors the same type of  hidden magnetic order\cite{jpr_10}.
Recent calculations by one of the authors also find electron-type Fermi surface pockets
at the corner of the two-iron Brillouin zone 
when electrons are added to the local moments\cite{jpr_17}.
Furthermore, in the previous section, we have pointed out that the low-energy
spin excitations predicted by the local-moment model in the hidden magnetic order phase
resemble the ``floating'' ring of spin-excitations
that has been observed recently in heavily electron-doped
FeSe by inelastic neutron scattering\cite{davies_16,pan_17,ma_17}.
The extended two-orbital Hubbard model therefore is promising phenomenologically.

Yet does it harbor superconductivity?
Recent exact calculations
by one of the authors
on finite clusters at the local-moment limit
find evidence for Cooper pairs near a quantum critical point 
to ``stripe'' spin-density wave order\cite{jpr_17}.
Also, quantum Monte Carlo simulations of a spin-fluctuation-exchange model
that is free of the sign problem,
and that is very similar to the model studied here
[Eqs. (\ref{hop}), (\ref{D}), and (\ref{e-hsw})],
find evidence for competition between SDW and superconducting groundstates\cite{Sachdev_12}.
(See also refs. \cite{Berg_16} and \cite{Lee_16}.)
It remains to be seen if the extended Hubbard model introduced here
also harbors superconductivity away from half filling, and if so, of what type.

\ack
The authors would like to thank Alexander Kass and Rong Yu for useful discussions.
This work was supported in part by the US Air Force
Office of Scientific Research under grant no. FA9550-17-1-0312
and by the National Science Foundation under PREM grant no. DMR-1523588.

\clearpage

\appendix

\section{Particle-Hole Symmetry}\label{ppndx_p_h}
Let us turn off next-nearest neighbor intra-orbital hopping (\ref{hop}), $t_2^{\parallel}=0$.
Consider then the following particle-hole transformation:
\begin{eqnarray}
c_{i,d\pm,s} \rightarrow e^{i{\bi Q}_{\rm AF}\cdot {\bi r}_i} c_{i,d\mp,s}^{\dagger} , \nonumber \\
c_{i,d\pm,s}^{\dagger} \rightarrow e^{-i{\bi Q}_{\rm AF}\cdot {\bi r}_i} c_{i,d\mp,s} ,
\label{p_h}
\end{eqnarray}
where ${\bi Q}_{\rm AF} = (\pi/a,\pi/a)$.
Making the above replacements in the Hamiltonian $H_{\rm hop}+H_U+H_{\rm sprx}$
for the extended two-orbital Hubbard model over the square lattice
(\ref{hop},\ref{U},\ref{sprx})
then results in the same Hamiltonian back up to a constant energy shift
and up to a shift in the chemical potential.  
The extended two-orbital Hubbard model
$H_{\rm hop}+H_U+H_{\rm sprx}$
is therefore symmetric under the particle-hole transformation (\ref{p_h})
at $t_2^{\parallel}=0$.

Next, substitution of the above particle-hole transformation (\ref{p_h})
into the creation operator for band electrons (\ref{ck})
yields the equivalent transformation in momentum space:
\begin{eqnarray}
c_{s} (n,{\bi k}) \rightarrow \pm i\, c_{s}^{\dagger}({\bar n},{\bar{\bi k}}) , \nonumber \\
c_{s}^{\dagger}(n,{\bi k}) \rightarrow \mp i\, c_{s}({\bar n},{\bar{\bi k}}) ,
\label{P_H}
\end{eqnarray}
where ${\bar n} = 1 + (n\; {\rm mod}\; 2)$, and where ${\bar{\bi k}} = {\bi k}+{\bi Q}_{\rm AF}$.
Here, we have used the property
\begin{equation}
\delta({\bi k} + {\bi Q}_{\rm AF}) = \pm {\pi\over 2} - \delta({\bi k})
\label{delta_k_pls_Q}
\end{equation}
satisfied by the phase shift, which is a result of the property
$\varepsilon_{\perp}({\bi k} + {\bi Q}_{\rm AF}) = - \varepsilon_{\perp}^*({\bi k})$
satisfied by the matrix element (\ref{mtrx_lmnt_b}). 
The hopping Hamiltonian (\ref{hop}) is
expressed in momentum space as
$$H_{\rm hop} = \sum_{\bi k} \sum_n \sum_s 
\varepsilon_n({\bi k}) c_s^{\dagger}(n,{\bi k}) c_s (n,{\bi k}).$$
It is invariant under the particle-hole transformation (\ref{P_H})
if the perfect nesting condition (\ref{prfct_nstng}) at $t_2^{\parallel} = 0$ holds true.
Here, we have used the property $\sum_{\bi k}\sum_n \varepsilon_n({\bi k}) = 0$.


\section{Antiferromagnetic magnetization and matrix elements}\label{ppndx_a}
Substituting the identity (\ref{ci}) for the creation operator
into expression (\ref{Sz}) for the antiferromagnetic magnetization,
along with the conjugate expression for the destruction operator,
yields the form 
\begin{eqnarray}
S_z (m,{\bi Q}_{\rm AF}) = {1\over 2}\sum_s\sum_{\bi k}\sum_{n,n^{\prime}} ({\rm sgn}\, s) {\cal M}_{n,{\bi k};n^{\prime},{\bar{\bi k}}} c_s^{\dagger}(n^{\prime},{\bar{\bi k}}) c_s(n,{\bi k}),\nonumber\\
\label{A_sz}
\end{eqnarray}
with ${\bar{\bi k}} = {\bi k} + {\bi Q}_{\rm AF}$,
and with matrix element
\begin{equation}
{\cal M}_{n,{\bi k};n^{\prime},{\bi k}^{\prime}} = {1\over 2} \sum_{\alpha=0,1}
e^{i (2 \alpha -1)[\delta({\bi k}) -\delta({\bi k}^{\prime})]} (-1)^{(n^{\prime}-n+m)\alpha}.
\label{A_M}
\end{equation}
The matrix element therefore equals
\begin{equation}
{\cal M}_{n,{\bi k};n^{\prime},{\bi k}^{\prime}} = 
\cases{\cos[\delta({\bi k})-\delta({\bi k}^{\prime})] & for $n^{\prime} = n + m\; ({\rm mod}\; 2)$,\\
-i\, \sin[\delta({\bi k})-\delta({\bi k}^{\prime})] & for $n^{\prime} = n + m + 1\; ({\rm mod}\; 2)$.}
\label{A_MM}
\end{equation}
Now replace ${\bi k^{\prime}}$ above with 
${\bar{\bi k^{\prime}}} = {\bi k^{\prime}} + {\bi Q}_{\rm AF}$,
and recall the definition of the phase shift: 
$e^{i 2 \delta} = \varepsilon_{\perp} / |\varepsilon_{\perp}|$.
Inspection of (\ref{mtrx_lmnt_b}) yields the identity
$\varepsilon_{\perp}({\bar {\bi k}}) = - \varepsilon_{\perp}^*({\bi k})$,
which in turn yields the identity (\ref{delta_k_pls_Q}).
It implies that
$\delta({\bi k}) - \delta({\bar{\bi k^{\prime}}}) = \delta({\bi k}) + \delta({\bi k^{\prime}})\mp \pi/2$.
Substituting this into the previous result (\ref{A_MM}) for the matrix element yields the final result
\begin{equation}
{\cal M}_{n,{\bi k};n^{\prime},{\bar{\bi k^{\prime}}}} = 
\cases{\pm\sin[\delta({\bi k})+\delta({\bi k^{\prime}})] & for $n^{\prime} = n + m\; ({\rm mod}\; 2)$,\\
\pm i\, \cos[\delta({\bi k})+\delta({\bi k^{\prime}})] & for $n^{\prime} = n + m + 1\; ({\rm mod}\; 2)$.}
\label{A_MMMM}
\end{equation}

\section{Transverse spin susceptibility}\label{ppndx_b}
Let us add a term 
$-h\sum_i\sum_{\alpha} S_{i,\alpha}^{(y)}$
to the mean-field Hamiltonian (\ref{HMF}) in the text.
Here, $h$ represents an external magnetic field applied along
the $y$ axis that is perpendicular to the sub-lattice magnetization of
the hidden antiferromagnet, which points along the $z$ axis.
Following the discussion in section \ref{cllctv_mds},
we shall quantize spin along the $y$-axis instead:
$z^{\prime} = y$, $x^{\prime} = z$ and $y^{\prime} = x$.
The mean-field Hamiltonian (\ref{HMF}) plus the additional terms above
then becomes
\begin{equation}
H^{(mf)} = \sum_s \sum_{\bi k}
\left[ {\begin{array}{c}
c_{s}^{\prime} \\ {\bar c}_{\bar s}^{\prime}
\end{array} } \right]^{\dagger}
\left[ {\begin{array}{cc}
\varepsilon_{+}-({\rm sgn}\, s){1\over 2}h    & \Delta   \\
\Delta  & -\varepsilon_{+}+({\rm sgn}\, s){1\over 2}h
\end{array} } \right]
\left[ {\begin{array}{c}
c_{s}^{\prime} \\ {\bar c}_{\bar s}^{\prime}
\end{array} } \right]
\label{Hmfprm}
\end{equation}
where $c_s^{\prime} ({\bi k}) = c_s^{\prime}(2,{\bi k})$, and 
where ${\bar c}_s^{\prime}({\bi k}) = c_s^{\prime}(1,{\bi k}+{\bi Q}_{\rm AF})$.
Above, ${\bar s} = -s$.
The energy eigenvalues of the quasi-particle excitations are then
\begin{equation}
E_{s}({\bi k}) = \sqrt{[\varepsilon_+({\bi k}) - ({\rm sgn}\, s) {1\over 2}h]^2+\Delta^2({\bi k})},
\label{E_s}
\end{equation}
with $\Delta({\bi k}) = \Delta_0 [\sin\, 2 \delta({\bi k})]$.
The transverse magnetization per iron atom is then
\begin{eqnarray}
M_y &=& 
{1\over 2}(a^2 N_{\rm Fe})^{-1} \sum_s\sum_{\bi k}\sum_n
 ({\rm sgn}\, s)\langle c_s^{\prime\dagger}(n,{\bi k}) c_s^{\prime}(n,{\bi k})\rangle \nonumber \\
&=& {1\over 2} (a^2 N_{\rm Fe})^{-1} \sum_{\bi k}\sum_{s}  ({\rm sgn}\, s) [v_s^2({\bi k}) - u_s^2({\bi k})],
\label{}
\end{eqnarray}
where  $u_{s}^2 ={1\over 2} + {1\over 2}[\varepsilon_+ -({\rm sgn}\, s){1\over 2}h]/E_{s}$
and $v_{s}^2 ={1\over 2} - {1\over 2}[\varepsilon_+ -({\rm sgn}\, s){1\over 2}h]/E_{s}$.
Substituting the latter in above then yields
\begin{equation}
M_y =
{1\over 2} (a^2 N_{\rm Fe})^{-1} \sum_{\bi k}
\Biggl(-{\varepsilon_+ - {1\over 2} h\over{E_{\uparrow}}}+{\varepsilon_+ + {1\over 2} h\over{E_{\downarrow}}}\Biggr).
\label{}
\end{equation}
Finally, taking the limit $h\rightarrow 0$ above results in the linear response
$M_y = \chi_{\perp}^{(0)} h$, with transverse spin susceptibility
\begin{equation}
\chi_{\perp}^{(0)}  = {1\over 2} (a^2 N_{\rm Fe})^{-1} \sum_{\bi k} {\Delta^2\over E^3}.
\label{}
\end{equation}

We will next prove that the integrals $I_1$ and $I_2$ 
defined by (\ref{intrmdt_a}) and (\ref{intrmdt_b}) 
can only be equal in the limit $t_2^{\perp}\rightarrow 0$ as $U_0\rightarrow\infty$.
First, observe that inspection of (\ref{s_2dlt}) yields the limit
${\rm lim}_{t_2^{\perp}\rightarrow 0} \sin\, 2\delta({\bi k}) = 0$.
Next, observe by (\ref{intrmdt_a}) that the limit
${\rm lim}_{U_0\rightarrow\infty,t_2^{\perp}\rightarrow 0} I_1$
is equal to
\begin{equation}
{\rm lim}_{U_0\rightarrow\infty,t_2^{\perp}\rightarrow 0} 
{1\over 2} N_{\rm Fe}^{-1}\sum_{\bi k} 
([\varepsilon_+({\bi k})/U(\pi)]^2+[\langle m_{0,0}\rangle \sin\,2\delta({\bi k})]^2)^{-1/2}.\nonumber
\label{i2}
\end{equation}
Figure \ref{magnetic_moment} indicates that 
the hidden magnetic moment $\langle m_{0,0}\rangle$ vanishes roughly
as the hybridization between the $3d_{xz}$ and $3d_{yz}$ orbitals, $|t_2^{\perp}|$.
We conclude that the limit
${\rm lim}_{U_0\rightarrow\infty,t_2^{\perp}\rightarrow 0} I_1$
diverges at least linearly with $U_0$.
Second, observe that the quotient in expression (\ref{intrmdt_b}) for $I_2$ is equal to
$2\,\delta(\Delta_0 [\sin\, 2\delta({\bi k})])$ in the limit $U_0\rightarrow\infty$.
By (\ref{Delta0}),
this yields the limiting expression
\begin{equation}
{\rm lim}_{U_0\rightarrow\infty}I_2 = a^2\int_{-\pi/a}^{+\pi/a} {d k_x\over{2\pi}}
\int_{-\pi/a}^{+\pi/a} {d k_y\over{2\pi}}
\delta(\langle m_{0,0}\rangle [\sin\, 2\delta({\bi k})]),
\end{equation}
which coincides with the product of $a^2 U(\pi)$ with 
the density of states of $\Delta_0 [\sin\, 2\delta({\bi k})]$ at zero energy.  
Now notice by (\ref{s_2dlt}) that 
$\sin\, 2\delta({\bi k})$ disperses hyperbolically near $(\pi/a,0)$ and $(0,\pi/a)$.
This implies that ${\rm lim}_{U_0\rightarrow\infty,t_2^{\perp}\rightarrow 0} I_2$ diverges roughly as
$(|t_1^{\perp}|/|t_2^{\perp}|)^2{\rm ln}[U(\pi)/W_{\rm bottom}]$.
Equating $I_1$ with $I_2$ then yields that $t_2^{\perp}\rightarrow 0$ as $U_0\rightarrow\infty$,
which is consistent with the original assumption.


\section{Eliashberg equations at non-zero temperature}\label{ppndx_c}
Equation (\ref{E_eqs_T}) in the text lists the three Eliashberg equations
at non-zero temperature in terms of sums over Matsubara frequencies.  
The sums can be evaluated in closed form
after a series of decompositions into partial fractions.
That procedure yields 
\begin{eqnarray}
[Z({\bi k},\omega)-1] \omega & = &
 \int {d^2  k^{\prime}\over{(2\pi)^2}}
{U^2(\pi)\over 2} {(2 s_1)^2\over{\chi_{\perp}}}
{\sin^2[\delta({\bi k})+\delta({\bi k}^{\prime})]
\over{2 Z({\bi k}^{\prime},\omega) \cdot 2\omega_b({\bi q})}}  \cdot \nonumber\\
&& \ \Biggl\{(n_{\rm F}[-E({\bi k}^{\prime})]+n_{\rm B}[\omega_b({\bi q})])\cdot\nonumber\\
&& \ \ \ \cdot
\Biggl[{1\over{\omega_b({\bi q})+E({\bi k}^{\prime})-\omega}}
-{1\over{\omega_b({\bi q})+E({\bi k}^{\prime})+\omega}}\Biggr]+\nonumber\\ 
&& \ \ \ (n_{\rm F}[+E({\bi k}^{\prime})]+n_{\rm B}[\omega_b({\bi q})])\cdot\nonumber\\
&& \ \ \ \cdot
\Biggl[{1\over{\omega_b({\bi q})-E({\bi k}^{\prime})-\omega}}
-{1\over{\omega_b({\bi q})-E({\bi k}^{\prime})+\omega}}\Biggr]\Biggr\},\nonumber\\
%
&&\\
%
%
%
-\nu & = &
\int {d^2 k^{\prime}\over{(2\pi)^2}}
{U^2(\pi)\over 2} {(2 s_1)^2\over{\chi_{\perp}}}
{\sin^2[\delta({\bi k})+\delta({\bi k}^{\prime})]
\over{2 Z({\bi k}^{\prime},\omega) \cdot 2\omega_b({\bi q})}}
{\varepsilon_+({\bi k}^{\prime})-\nu\over{Z({\bi k}^{\prime},\omega) E({\bi k}^{\prime})}}  \cdot \nonumber\\
&& \ \Biggl\{(n_{\rm F}[-E({\bi k}^{\prime})]+n_{\rm B}[\omega_b({\bi q})])\cdot\nonumber\\
&& \ \ \ \cdot
\Biggl[{1\over{\omega_b({\bi q})+E({\bi k}^{\prime})-\omega}}
+{1\over{\omega_b({\bi q})+E({\bi k}^{\prime})+\omega}}\Biggr]- \nonumber\\
&& \ \ \ (n_{\rm F}[+E({\bi k}^{\prime})]+n_{\rm B}[\omega_b({\bi q})])\cdot\nonumber\\
&& \ \ \ \cdot
\Biggl[{1\over{\omega_b({\bi q})-E({\bi k}^{\prime})-\omega}}
+{1\over{\omega_b({\bi q})-E({\bi k}^{\prime})+\omega}}\Biggr]\Biggr\}, \nonumber\\
%
&&\\
%
Z({\bi k},\omega) \Delta({\bi k}) & = &
 \int {d^2 k^{\prime}\over{(2\pi)^2}}
{U^2(\pi)\over 2} {(2 s_1)^2\over{\chi_{\perp}}}
{\sin^2[\delta({\bi k})+\delta({\bi k}^{\prime})]
\over{2 Z({\bi k}^{\prime},\omega) \cdot 2\omega_b({\bi q})}}
{\Delta({\bi k}^{\prime})\over{E({\bi k}^{\prime})}} \cdot \nonumber\\
&&\ \ \ \ \Biggl\{(n_{\rm F}[-E({\bi k}^{\prime})]+n_{\rm B}[\omega_b({\bi q})])\cdot\nonumber\\
&& \ \ \ \ \ \ \ \cdot
\Biggl[{1\over{\omega_b({\bi q})+E({\bi k}^{\prime})-\omega}}
+{1\over{\omega_b({\bi q})+E({\bi k}^{\prime})+\omega}}\Biggr]- \nonumber\\
&& \ \ \ \ \ \ (n_{\rm F}[+E({\bi k}^{\prime})]+n_{\rm B}[\omega_b({\bi q})])\cdot\nonumber\\
&& \ \ \ \ \ \cdot
\Biggl[{1\over{\omega_b({\bi q})-E({\bi k}^{\prime})-\omega}}
+{1\over{\omega_b({\bi q})-E({\bi k}^{\prime})+\omega}}\Biggr]\Biggr\}. \nonumber\\
\label{3_E_eqs}
\end{eqnarray}
Above, ${\bi q} = {\bi k} - {\bi k}^{\prime} - {\bi Q}_{\rm AF}$.  
Also, $n_{\rm F}(\varepsilon)$ and $n_{\rm B}(\omega)$
denote the Fermi-Dirac and the Bose-Einstein distribution functions.

\end{document}